\documentclass[aps,prb,superscriptaddress,twocolumn,showpacs,amsmath,amssymb,]{revtex4}
\usepackage{array}
\usepackage{graphicx}
\usepackage{dcolumn}
\usepackage{bm}
\usepackage[latin1]{inputenc}
\usepackage{graphicx,color,epsfig,rotate}

\newcommand{\mb}{$\mu_{\text{B}}$}

\begin{document}

\title{Interplay of rare earth and iron magnetism in \textit{R}FeAsO with \textit{R} = La, Ce, Pr, and Sm: A muon spin relaxation study and symmetry analysis}


\author{H. Maeter}
\affiliation{Institut f\"ur Festk\"orperphysik, TU Dresden, D--01069 Dresden, Germany}
\author{H. Luetkens}
\affiliation{Laboratory for Muon--Spin Spectroscopy, Paul Scherrer Institut, CH--5232 Villigen PSI, Switzerland}
\author{Yu. G. Pashkevich}
\affiliation{A. A. Galkin Donetsk Phystech NASU, 83114 Donetsk, Ukraine}
\author{A. Kwadrin}
\affiliation{Institut f\"ur Festk\"orperphysik, TU Dresden, D--01069 Dresden, Germany}
\author{R. Khasanov}
\affiliation{Laboratory for Muon--Spin Spectroscopy, Paul Scherrer Institut, CH--5232 Villigen PSI, Switzerland}
\author{A. Amato}
\affiliation{Laboratory for Muon--Spin Spectroscopy, Paul Scherrer Institut, CH--5232 Villigen PSI, Switzerland}
\author{A. A. Gusev}
\affiliation{A. A. Galkin Donetsk Phystech NASU, 83114 Donetsk, Ukraine}
\author{K. V. Lamonova}
\affiliation{A. A. Galkin Donetsk Phystech NASU, 83114 Donetsk, Ukraine}
\author{D. A. Chervinskii}
\affiliation{A. A. Galkin Donetsk Phystech NASU, 83114 Donetsk, Ukraine}
\author{R. Klingeler}
\affiliation{Leibniz--Institut f\"ur Festk\"orper-- und Werkstoffforschung (IFW) Dresden, D--01171 Dresden,
Germany}
\author{C. Hess}
\affiliation{Leibniz--Institut f\"ur Festk\"orper-- und Werkstoffforschung (IFW) Dresden, D--01171 Dresden,
Germany}
\author{G. Behr}
\affiliation{Leibniz--Institut f\"ur Festk\"orper-- und Werkstoffforschung (IFW) Dresden, D--01171 Dresden,
Germany}
\author{B. B\"uchner}
\affiliation{Leibniz--Institut f\"ur Festk\"orper-- und Werkstoffforschung (IFW) Dresden, D--01171 Dresden,
Germany}
\author{H.--H. Klauss}
\email{h.klauss@physik.tu-dresden.de}
\affiliation{Institut f\"ur
Festk\"orperphysik, TU Dresden, D--01069 Dresden, Germany}


\date{\today}

\begin{abstract}
We report zero field muon spin relaxation ($\mu$SR) measurements on
\textit{R}FeAsO with \textit{R} = La, Ce, Pr, and Sm. We study the
interaction of the FeAs and \textit{R} (rare earth)  electronic
systems in the non superconducting magnetically ordered parent
compounds of \textit{R}FeAsO$_{1-x}$F$_x$ superconductors via a
detailed comparison of the local hyperfine fields at the muon site
with available Mössbauer spectroscopy and neutron scattering data.
These studies provide microscopic evidence of long range
commensurate magnetic Fe order with the Fe moments not varying by
more than 15~\% within the series \textit{R}FeAsO with \textit{R} =
La, Ce, Pr, and Sm. At low temperatures, long range \textit{R}
magnetic order is also observed. Different combined Fe and
\textit{R} magnetic structures are proposed for all compounds using
the muon site in the crystal structure obtained by electronic
potential calculations. Our data point to a strong effect of
\textit{R} order on the iron subsystem in the case of different
symmetry of Fe and \textit{R} order parameters resulting in a Fe
spin reorientation in the \textit{R} ordered phase in PrFeAsO. Our
symmetry analysis proves the absence of collinear Fe--\textit{R}
Heisenberg interactions in \textit{R}FeAsO. A strong Fe--Ce coupling
due to non--Heisenberg anisotropic exchange is found in CeFeAsO
which results in a large staggered Ce magnetization induced by the
magnetically ordered Fe sublattice far above $T_N^{Ce}$. Finally, we
argue that the magnetic \textit{R}--Fe interaction is probably not
crucial for the observed enhanced superconductivity in
\textit{R}FeAsO$_{1-x}$F$_x$ with a magnetic \textit{R} ion.
\end{abstract}

\pacs{75.30.Fv, 74.70.-b, 76.75.+i, 76.80.+y}


\maketitle


\section{Introduction}
The recent discovery of high temperature superconductivity in
LaFeAsO$_{1-x}$F$_x$ by Kamihara and coworkers has triggered intense
research in the Fe--pnictides.\cite{Kamihara08} In its wake,
superconductivity with transition temperatures that exceed 50~K have
been found in the oxopnictide materials in which La is substituted
by \textit{R} = Sm, Ce, Nd, Pr, and Gd, respectively.
\cite{Chen08_XH-arXiv,Chen08_GFb-arXiv,Ren08a-arXiv,Ren08b-arXiv,Cheng08-arXiv}
Besides the high critical temperatures, striking similarities to the
properties of the high--T$\rm_c$ cuprates have been pointed out. As
with the cuprates, the Fe--pnictides have a layered crystal
structure with alternating FeAs and \textit{R}O sheets, where the Fe
ions are arranged on a simple square lattice.\cite{Kamihara08}
Superconductivity emerges in the pnictides when doping the antiferromagnetic parent compound either with electrons or holes which suppresses the magnetic order.\cite{Luetkens08a,Luetkens08b-arXiv} These similarities raised the hope that cuprates and pnictides share a common mechanism for superconductivity, and that after 20 year of research on high--T$\rm_c$ cuprates the Fe--pnictides may provide new insight into the superconducting coupling mechanism and verify existing theories about high temperature superconductivity.

In contrast to the cuprates, the non superconducting magnetic parent
compound is not a Mott--Hubbard insulator but a metal. Theoretical
studies reveal a two--dimensional electronic structure with all Fe
3d bands contributing to the density of states at the Fermi
level.\cite{Singh08-arXiv} Neutron diffraction\cite{Cruz08} and
local probe techniques like Mössbauer spectroscopy and
$\mu$SR\cite{Klauss08,Carlo09, Bernhard09} prove that the
\textit{R}FeAsO parent compounds order in a commensurate spin
density wave (SDW) magnetic order with a strongly reduced ordered Fe
moment. Neutron studies suggest a columnar magnetic structure with a
Fe magnetic moment between 0.25~\mb{} and 0.8~\mb{} below
$T_{N}\approx 140$~K that depends on \textit{R}.\cite{Lynn08} Due to
the small size of the ordered SDW moment (compared to metallic iron
with a moment of approx. 2.2~\mb{} per Fe) and the lack of large
single crystals the temperature dependence of the magnetic order
parameter can be determined by local probe techniques with a much
higher accuracy than with neutron scattering.\cite{Klauss08} Note
that the magnetic transition in the \textit{R}FeAsO system is always
preceeded by an orthorhombic structural distortion which appears at
$T_{S}$, which is about 10--20~K above $T_{N}$.

It is still an open question why the rare earth containing systems
in the series \textit{R}FeAsO$_{1-x}$F$_x$ with a localized
\textit{R} magnetic moment have a higher T$\rm_c$ than
LaFeAsO$_{1-x}$F$_x$. One suggestion is based on a purely
geometrical argument. It is argued that the different ionic radii of
the rare earth elements change the Fe--As--Fe bond angles in the
Fe--As plane and the FeAs--FeAs interplane distance. As a
consequence, the planar anisotropy of the electronic properties may
be better for superconductivity in case of \textit{R} = Ce, Pr, Nd,
etc.. On the other hand, the electronic interaction of the
\textit{R} 4f electrons with the Fe conduction band states may be
crucial to enhance the density of states at the Fermi energy.

In this work we examine the interaction of the FeAs electronic bands
with the rare earth subsystem by a detailed comparison of the local
hyperfine fields at the muon site and Fe nucleus with neutron
scattering results. These studies were performed on the undoped
magnetically ordered parent compounds of the
\textit{R}FeAsO$_{1-x}$F$_x$ superconductors. We report zero field
muon spin relaxation measurements on powder samples of
\textit{R}FeAsO with \textit{R} = La, Ce, Pr, and Sm. We provide
microscopic evidence of static commensurate magnetic order of Fe
moments. The N\'eel temperatures and the temperature dependence of
the Fe sublattice magnetization were determined with high precision
and are compared with available Mössbauer spectroscopy and neutron
scattering results from the literature. In contrast to neutron
studies our measurements prove that the size of the ordered Fe
moment is independent of the rare earth ion. For a quantitative
analysis of the $\mu$SR spectra the muon site in the \textit{R}FeAsO
crystal structure has been determined by electronic potential
calculations using a modified Thomas--Fermi approach. Spontaneous
magnetic ordering of the rare earth magnetic moment is observed by
$\mu$SR below $T_N^{R}=$~4.4(3), 11(1), and 4.66(5)~K for \textit{R}
= Ce, Pr, and Sm, respectively. Iron and \textit{R} magnetic
structures are proposed for all compounds on the basis of a detailed
symmetry analysis and magnetic dipole field calculations at the muon
site on the one hand and $\mu$SR, Mössbauer spectroscopy and neutron
scattering data on the other. Non--collinear magnetic order of Ce
and Sm is found in the corresponding compounds, which we explain by
a weak coupling of adjacent \textit{R} planes in the
\textit{R}--O--\textit{R} layer. In PrFeAsO the $\mu$SR data suggest
an Fe spin reorientation developing below $T_{N}^{Pr}$, while the Fe
order is unaltered below $T_{N}^{\textit{R}}$ in the Sm and Ce
compounds.

In CeFeAsO we find a sizable staggered magnetization of the Ce ions
induced by the Fe subsystem already far above $T^{Ce}_N$ which
amounts to approximately $0.3$~\mb{} near to $T^{Ce}_N$. We argue
that the neglect of the Ce magnetization may have caused the
overestimation of the ordered Fe moment in recent neutron
diffraction studies. Our symmetry analysis proves the absence of
collinear Fe--\textit{R} Heisenberg interaction in \textit{R}FeAsO
compounds. Using classical and quantum mechanical approaches we
deduce Fe--Ce and Ce--Ce exchange coupling constants and show that
the Fe--Ce non--Heisenberg exchange interaction is exceptionally
strong and of the same order as the Ce--Ce exchange interaction. In
the Sm and Pr compounds the observed coupling between the \textit{R}
and the Fe subsystems is found to be much weaker than in CeFeAsO.
Therefore, we conclude that the magnetic \textit{R}--Fe interaction
is probably not crucial for the enhanced superconducting transition
temperatures in \textit{R}FeAsO$_{1-x}$F$_x$ with magnetic 4f ions
compared to LaFeAsO$_{1-x}$F$_x$, since only in CeFeAsO a strong
\textit{R}--Fe coupling is observed.

\section{Experimental}
\label{sec.exp} Polycrystalline \textit{R}FeAsO have been prepared
by using a two--step solid state reaction method, similar to that
described by Zhu et al., and annealed in vacuum.\cite{Zhu_X08-arXiv}
The crystal structure and the composition were investigated by
powder X--ray diffraction. From the X--ray diffraction data no
impurity phases are inferred.

In a $\mu$SR experiment nearly 100\% spin--polarized muons are implanted into the sample one at a time. The
positively charged $\mu^+$ thermalize at interstitial lattice sites, where they act as magnetic micro probes. In
a magnetic material the muon spin precesses about the local magnetic field $B$ at the muon site with the Larmor
frequency $f_{\mu} = \gamma_\mu/(2\pi) B$ (muon gyromagnetic ratio $\gamma_\mu /2 \pi = 135.5$
MHz~T$^{-1}$). With a lifetime of $\tau_\mu = 2.2$~$\mu$s the muon decays into two neutrinos and a positron,
the latter being predominantly emitted along the direction of the muon spin at the moment of the decay.
Measurement of both the direction of positron emission as well as the time between muon implantation and
positron detection for an ensemble of several millions of muons provides the time evolution of the muon spin
polarization $P(t)$ along the initial muon spin direction. Magnetic materials with commensurate order possess a
well--defined local field $B$ at the muon site. Therefore, a coherent muon spin precession can be observed, which
for powder samples has the following functional form (see e.g. Ref.~\onlinecite{Reotier97}):

\begin{equation}
 P(t) = \sum_{i=1}^n P_i \left[ \frac{2}{3} e^{-\lambda_T^i t} \cos(\gamma_\mu B_i t+\phi) + \frac{1}{3} e^{-\lambda_L^i t} \right]
 \label{eqn.relax}
\end{equation}

\noindent The occurrence of 2/3 oscillating and 1/3 non--oscillating $\mu$SR signal fractions originates from
the spatial averaging in powder samples where 2/3 of the magnetic field components are perpendicular to the
$\mu^+$ spin and cause a precession, while the 1/3 longitudinal field components do not. The relaxation of the
oscillation, $\lambda_T$, is a measure of the width of the static field distribution $\Delta
B=\lambda_T/\gamma_\mu$. Dynamical effects are also present in $\lambda_T$ while the relaxation of the second term, $\lambda_L$, is due to dynamic magnetic fluctuations only.

If $n$ magnetically inequivalent muon sites exist in the crystallographic or magnetic structure, each of the
sites contribute to the $\mu$SR signal with its weight $P_i$. In 100\% magnetically ordered specimens $\sum P_i
= 1$. Therefore $\mu$SR not only provides a highly sensitive measure of the magnetic order parameter via
internal magnetic fields $B$, but also allows to independently determine the magnetic volume fraction. This is not
possible with non--local probes such as e.g. neutron diffraction.

\section{Muon spin relaxation results}\label{sec.muon}
\begin{figure*}[htbp]
\includegraphics[width=17.9cm]{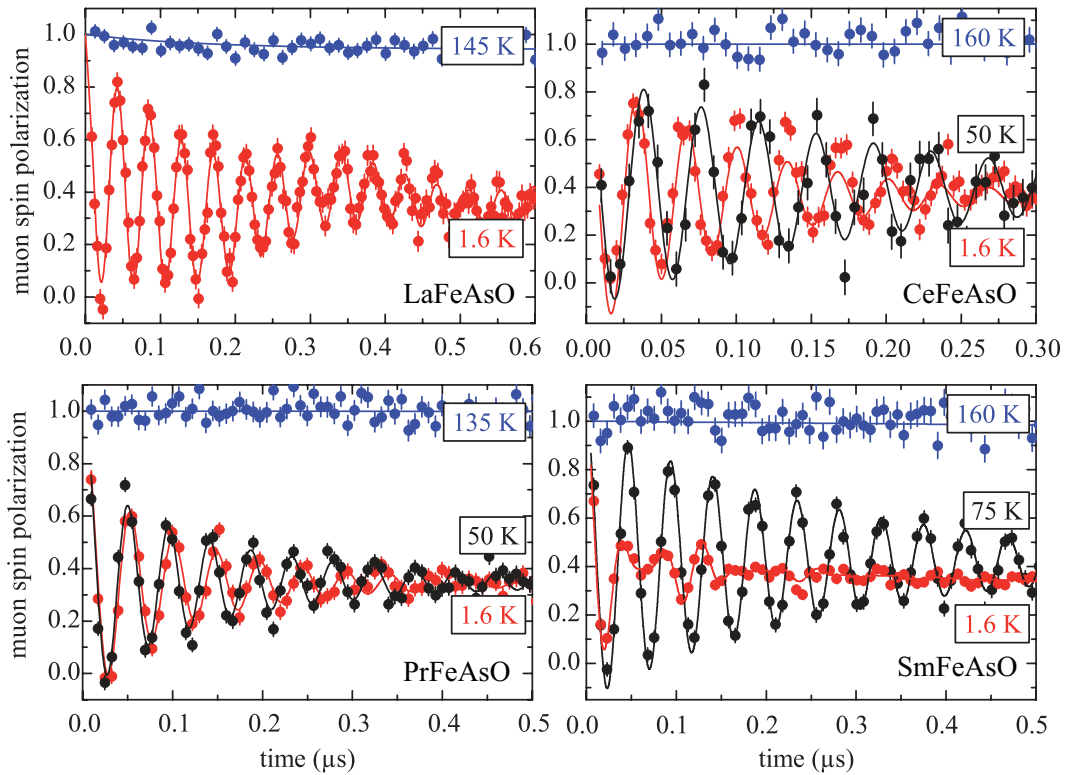}
\caption[]{Zero field $\mu$SR spectra for \textit{R}FeAsO with
\textit{R} = La, Ce, Pr, and Sm for three different temperatures:
$T>T_{N}$, $T_{N}>T>T_{N}^{R}$, and $T<T_{N}^{R}$.}
\label{img.spectra}
\end{figure*}

\begin{figure}[htbp]

\includegraphics[width=8.6cm]{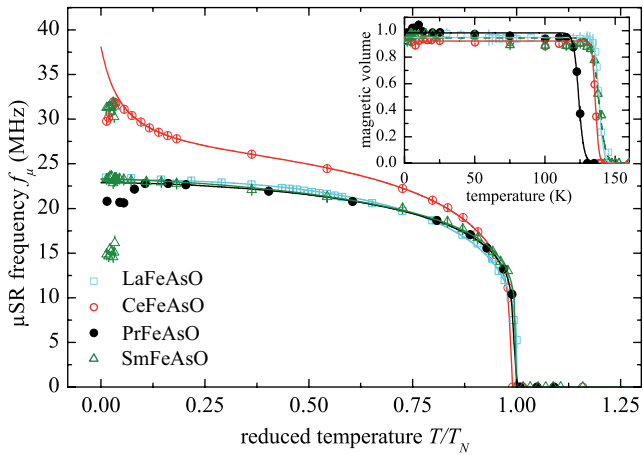}
\caption[]{Muon spin precession frequency as a function of reduced
temperature for \textit{R}FeAsO with \textit{R} = La, Ce, Pr, and
Sm. Inset: Magnetic signal fraction for \textit{R}FeAsO with
\textit{R} = La, Ce, Pr, and Sm. Lines are guides to the eye.}
\label{img.fraction}

\end{figure}

In Fig.~\ref{img.spectra}, zero field (ZF) $\mu$SR time spectra are
shown for \textit{R}FeAsO with \textit{R} = La, Ce, Pr, and Sm. At
high temperatures above 150~K no muon spin precession and only a
very weak depolarization of the $\mu^+$ polarization $P(t)$ is
observed. This weak depolarization and its Gaussian shape are
typical for a paramagnetic material and reflect the occurrence of a
small Gaussian--Kubo--Toyabe depolarization originating from the
interaction of the $\mu^+$ spin with randomly oriented nuclear
magnetic moments.\cite{Hayano79} At temperatures below $T_{N}$ a
well--defined spontaneous muon spin precession is observed in all
compounds indicating a well--defined magnetic field at the muon
sites. Therefore, long range static magnetic order with a
commensurate magnetic structure is realized in all investigated
compounds of the \textit{R}FeAsO series. Accordingly, incommensurate
order or spin glass magnetism can be excluded. Only in LaFeAsO, a
second $\mu$SR frequency with lower amplitude $P_2 \approx 0.30$ is
observed in addition to the main precession signal $P_1 \approx
0.70$. As described above, this indicates that two magnetically
inequivalent muon stopping sites are present in the crystal
lattice/magnetic structure. \label{sec.results}
The (static) magnetic order develops in the full sample volume below $T_{N}$ as evidenced by the magnetic $\mu$SR
signal fraction shown in the inset of Fig.~\ref{img.fraction}. The observed 5--10\% residual non--magnetic signal fraction observed
in our measurements are due to muons that do not hit the sample but stop in the sample holder or cryostat walls.

The N\'eel temperatures obtained from the $\mu$SR measurements
(superscript $\mu$) are summarized in table~\ref{tableTc}. As we
have shown previously for LaFeAsO, the N\'eel temperature $T_N$ of
the iron subsystem and the structural transition temperature $T_S$
can also be determined from anomalies in the temperature dependence
of the resistivity.\cite{Klauss08} For all \textit{R}FeAsO compounds
investigated here, pronounced anomalies have been
observed.\cite{Klauss08,Kimber08,Hess08-arXiv,Hamann-Borrero-unpublished}
The corresponding transition temperatures $T_{N}^\rho$ and
$T_{S}^\rho$ are listed also in table~\ref{tableTc}. In addition
magnetic, $T_{N}^{n}$ of the iron and the rare earth (superscript
$R$) subsystems, and structural, $T_{S}^{n/x}$, transition
temperatures deduced from neutron (superscript $n$) and X--ray
scattering (superscript $x$) experiments are given for comparison.
Note that for SmFeAsO no neutron data are available due to the high
neutron absorption of natural Sm.

\begin{table}[htbp]
\caption{Magnetic and structural transition temperatures of
\textit{R}FeAsO with \textit{R} = La, Pr, Sm, and Ce. In the cases
where no reference is given the transition temperatures were
measured exactly on the samples used in this
study.\cite{Klauss08,Luetkens08b-arXiv,Hess08-arXiv,Kimber08,Hamann-Borrero-unpublished}
All temperatures are given in Kelvin. See text for the abbreviations
in the superscripts.\label{tableTc}} \footnotesize\rm
\begin{ruledtabular}
\begin{tabular}{@{}*{7}{lccccccc}}
\textit{R} & $T_N^{\mu,R}$& $T_{N}^{n,R}$& $T_{N}^\mu $& $T_{N}^{n}$ & $T_{N}^\rho $& $T_{S}^{n/x} $& $T_{S}^\rho $\\
\hline
La&--&--&139(1)&137(3)\cite{Cruz08}&138.0&158(3)&156.0\\
Ce&4.4(3)&$\approx$5\cite{Zhao08a}&137(2)&139(5)\cite{Zhao08a}&134.8&152.2&151.5\\
Pr&11(1)&11&123(2)&127.0\cite{Zhao08b}&127.0&136.0&136.0\\
Sm&4.66(5)&--&138(2)&--&136.5&158.5&160.0\\
\end{tabular}
\end{ruledtabular}
\end{table}

Within the experimental error, the N\'eel temperatures determined from $\mu$SR, resistivity and neutron scattering agree very
well. The structural transition is found to be clearly separated by 10--20~K from $T_{N}$ for all investigated
compounds. Our highly sensitive $\mu$SR investigations, prove the absence of static long range magnetic order
between $T_{N}$ and $T_{S}$. Also no static magnetic short range correlations or disordered magnetism, which
would have been easily detected by $\mu$SR, has been observed. However, our $\mu$SR data do not rule out a
dynamic nematic magnetic phase with broken Ising symmetry that has recently been proposed to develop below
$T_{S}$,\cite{Xu08,Fang08} provided that the fluctuations are faster than approx. 10~GHz.

\subsection{Iron magnetic order}
\begin{figure*}[htbp]
\includegraphics[width=17.9cm]{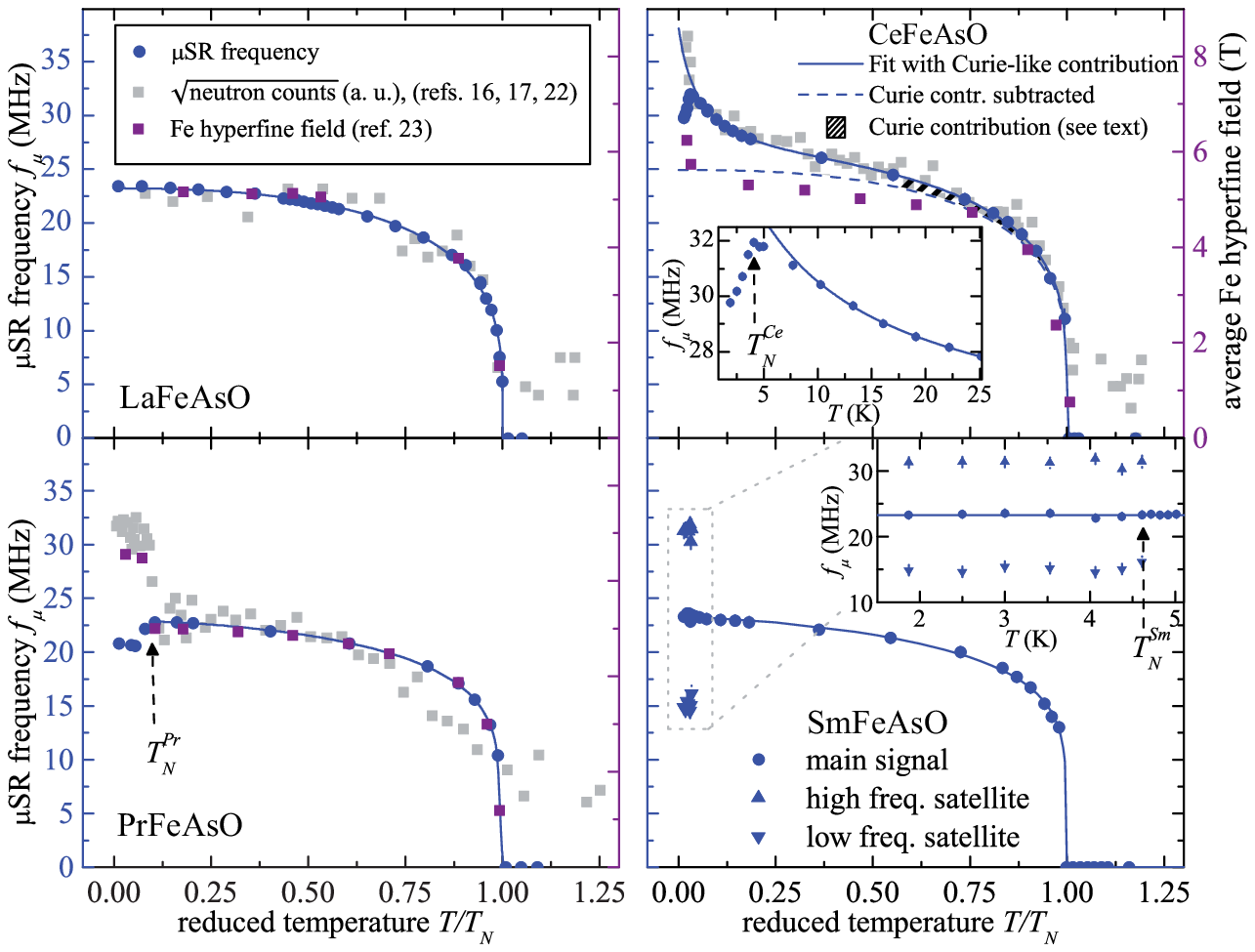}
\caption{Main $\mu$SR frequency as a function of reduced temperature $T/T^{Fe}_N$ together with the average magnetic hyperfine field at the Fe site from
Mössbauer spectroscopy \cite{McGuire08b-arXiv} and the square root of the magnetic Bragg peak intensity of
available neutron scattering \cite{Huang08,Zhao08a,Zhao08b} data. Note that the scale of $f_{\mu SR}$ and of
$B_{hf}$ is the same in all graphs. Some axes' labels have been omitted for clarity. Typically, error bars of the $\mu$SR frequency are smaller than
the data points.} \label{img.orderparameter}
\end{figure*}

Now we turn our discussion to the temperature dependence of the Fe
sublattice magnetization. We will first concentrate on the
temperature regime above the static order of the rare earth moment.
The temperature dependence of the muon spin precession frequency for
\textit{R}FeAsO with \textit{R} = La, Ce, Pr, and Sm is shown in
Fig.~\ref{img.fraction} for comparison. All compounds display a very
similar temperature dependence and absolute value of the $\mu$SR
frequency. Only the \textit{R} = Ce compound shows a higher
frequency and a stronger temperature dependence below $T_N$. In
Fig.~\ref{img.orderparameter} the $\mu$SR frequency $f_{\mu}(T)$ is
shown together with the average magnetic hyperfine field $B_{hf}(T)$
at the Fe site from $^{57}$Fe Mössbauer spectroscopy
\cite{McGuire08b-arXiv} and the square root of the magnetic Bragg
peak intensity $\sqrt{I(T)}$ of available neutron scattering
data.\cite{Huang08,Zhao08a,Zhao08b} The scale for $f_{\mu}(T)$ and
$B_{hf}(T)$ is the same for all diagrams, and the scales for
$\sqrt{I(T)}$ have been adjusted so that the $\mu$SR and neutron
data coincide as much as possible. In the following, we discuss the
data for each compound separately.

\subsubsection{LaFeAsO}

In LaFeAsO all three observables $f_{\mu}(T)$, $B_{hf}(T)$, and
$\sqrt{I(T)}$ scale with each other since they all measure the size
of the ordered Fe moment. A gradual increase of the ordered Fe
moment observed below $T_{N}$ indicates a second order phase
transition. The steep increase of the order parameter deviates from
the mean--field $\sqrt{1-\left(T/T_{N}\right)^2}$ temperature
dependence. As we have shown previously, this can be understood
qualitatively in the framework of a four band spin density
model.\cite{Klauss08} The onset of the magnetic order is accompanied
by a broad static magnetic field distribution $\Delta
B=\lambda_T/\gamma_\mu$ at the muon site (see Sec.~\ref{sec.exp})
that narrows rapidly with decreasing temperature, as can be seen in
Fig.~\ref{img.rate}(a). The relative width of the field distribution
$\Delta_{rel}=\Delta B / B$ is proportional to $(T-T^{Fe}_N)^{-1}$,
i.e. $\Delta_{rel}$ diverges with $T^{-1}$ as the temperature
approaches the N\'eel temperature of the Fe sublattice from below
for all systems presented here (not shown). Except for SmFeAsO no
dynamic magnetic fluctuations are detected ($\lambda_L=0$) for
temperatures below $T^{Fe}_N$.

\subsubsection{SmFeAsO}
\begin{figure}[htbp]
\includegraphics[width=8.65cm]{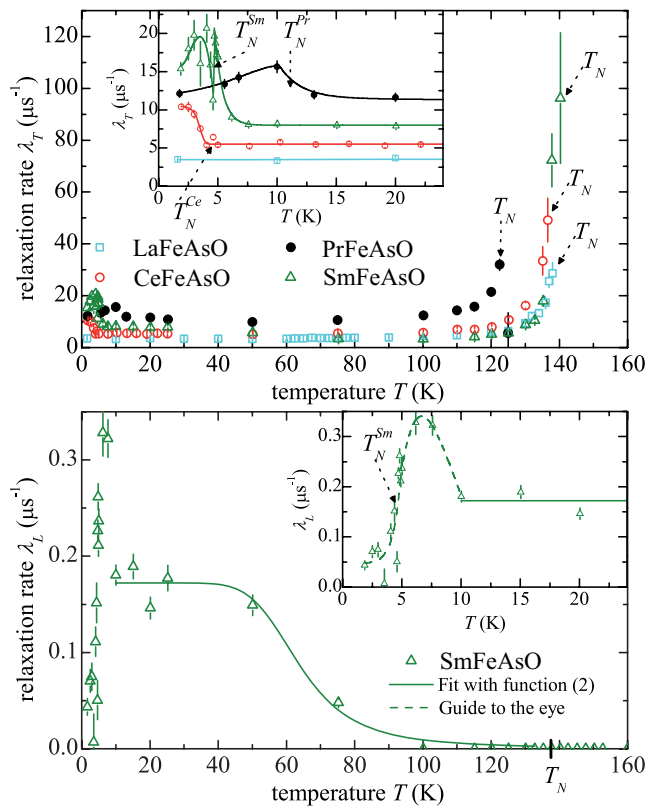}
\caption{\textbf{top}: Transverse relaxation rate $\lambda_T(T)$ for
\textit{R}FeAsO with \textit{R} = La, Ce, Pr, and Sm. Lines are
guides to the eye. \textbf{bottom}: Longitudinal relaxation rate
$\lambda_L(T)$ for \textit{R}FeAsO with \textit{R} = Sm only. The
solid line is a fit of equation \ref{eqn.ratel} to the data. The
dashed line is a guide to the eye} \label{img.rate}
\end{figure}

For SmFeAsO no neutron and Mössbauer data are available.
Qualitatively and quantitatively the temperature dependence of the
observed $\mu$SR frequency is very similar to that of LaFeAsO and
consistent with previously reported $\mu$SR data.\cite{DrewNature} A
sharp increase of the ordered Fe moment is observed below
$T_N^{Fe}$. Also the saturation value of $f_{0} \approx 23$~MHz
indicates the same size of the ordered Fe moment in LaFeAsO and
SmFeAsO assuming the same hyperfine coupling constants in these
isostructural compounds. In contrast to the other systems, magnetic
fluctuations in the time window of $\mu$SR are detected in this
system that cause the dynamic relaxation rate $\lambda_L$ to
increase gradually and to saturate between 10~K and 30~K. As
reported by Khasanov et al., this can be associated with fluctuating
Sm magnetic moments due to a thermally activated population of Sm
crystal electric field levels.\cite{Khasanov08} The temperature
dependence of the dynamic relaxation rate is well described by

\begin{equation}
\frac{1}{\lambda_L(T)}=\frac{1}{\lambda^0_L}+\frac{1}{C\exp(E_0/k_BT)},\label{eqn.ratel}
\end{equation}

\noindent with a saturation value $\lambda^0_L$ at low temperatures
and an activation energy $E_0$ that is related to low lying Sm
crystal electric field levels. The rough agreement of the activation
energy $E_0$ of the Sm magnetic moment fluctuations with the Sm
crystal electric field splitting has been confirmed by specific heat
measurements reported by Baker and coworkers.\cite{Baker08-arxiv} A
fit of Eq.~(\ref{eqn.ratel}) to the longitudinal relaxation rate
$\lambda_L(T)$ is shown in Fig.~\ref{img.rate} and yields
$\lambda^0_L=0.172(6)~\mu$s$^{-1}$, $C=0.001~\mu$s$^{-1}$, and
$E_0=44(10)$~meV. This value for $E_0$ is approximately two times
higher than in the oxygen deficient SmFeAsO$_{0.75}$ studied by
Khasanov et al..\cite{Khasanov08} However, one has to take into
account that the oxygen deficiency in SmFeAsO$_{0.75}$ causes a
higher defect density and changes of the lattice
parameters.\cite{Ren08}
Therefore, a different activation energy $E_0$ for the undoped SmFeAsO compared to SmFeAsO$_{0.75}$ is possible.

\subsubsection{PrFeAsO}
Apart from a slightly reduced N\'eel temperature, PrFeAsO shows the
same temperature dependence and saturation value of the Fe
sublattice magnetization, i.e. $f_{\mu}(T)$ as SmFeAsO. Note that
again the saturation value for $f_{\mu}$ and $B_{hf}$ are very
similar to that of LaFeAsO indicating the same size of the ordered
Fe moment. A close comparison of $f_\mu(T)$ and $B_{hf}(T)$ reveals
that $f_\mu(T)$ is systematically reduced by a small amount compared
to $B_{hf}(T)$. This phenomenon can be explained by the same
mechanism found in the Ce system (see below and
Section~\ref{sec.magn-structures}). In the case of PrFeAsO the muon
spin polarization function (\ref{eqn.relax}) did not approximate the
data very well and a non--zero phase $\phi$ of the oscillation and a
generalized exponential relaxation function $\exp(-(\lambda_T
t)^\alpha)$ had to be used to describe the data.

\subsubsection{CeFeAsO}
A qualitatively different behavior is observed for the CeFeAsO
compound. Neutron diffraction as well as $\mu$SR data do not scale
with the observed hyperfine field $B_{hf}$ at the Fe site. The
magnetic Bragg intensity as well as the internal magnetic field at
the muon site measured by the muon precession frequency continuously
increase below $T_{N}$. Only the Mössbauer hyperfine field displays
the same rapid saturation below $T_N$ with the same ordered moment
as observed for the La and Pr compounds.\cite{McGuire08b-arXiv}
Therefore we conclude that the $\mu$SR as well as the neutron data
do not solely measure the Fe sublattice magnetization, but also
contain a significant contribution from the Ce sublattice. The
$^{57}$Fe Mössbauer spectroscopy provides the most accurate
measurement of the on--site Fe sublattice magnetization without
sizable contribution from the rare earth moments due to a weak
transferred hyperfine coupling. In contrast, a magnetization on the
rare earth site induced by the Fe subsystem has the same symmetry as
the Fe order and therefore contributes to the same Bragg peaks. In
principle, neutron scattering can distinguish between the different
contributions from the Fe and the Ce sublattice by fitting the
different magnetic form-factors. This has not been done for
$T>T_N^{R}$ in the neutron studies \cite{Huang08,Zhao08a,Zhao08b}
shown in Fig.~\ref{img.orderparameter}, where the whole magnetic
intensity has been attributed to originate from Fe moments alone.
Similar to the neutron data, the local field at the muon site also
contains a contribution from the rare earth magnetic sublattice. In
the following, we model the above mentioned contribution of the
induced Ce moment to the field at the muon site by an additional
Curie--Weiss contribution. This can be interpreted as a local
magnetization of paramagnetic moments of the Ce sublattice induced
by the molecular field generated by the Fe sublattice. In turn, the
induced Ce sublattice magnetization creates a dipole field at the
muon site. The temperature dependence of the Fe molecular field has
to be proportional to the Fe sublattice magnetization, i.e. the muon
spin precession frequency $f_{\mu}(T)$ observed in the other
\textit{R}FeAsO systems. This is plausible because we have shown
that $f_{\mu}(T)$ above the rare earth ordering temperature
$T^{R}_{N}$ is almost independent of the rare earth ion. Thus we
chose the following function to describe our data in a first
approximation:

\begin{eqnarray}
f_{\mu}(T)&=&f_0
\left[1-\left(\frac{T}{T^{Fe}_N}\right)^{\alpha}\right]^{\beta}\cdot\left[1+\frac{\tilde{C}}{T-\theta}\right].
\label{eqn.ce}
\end{eqnarray}

\noindent The first term in the last square brackets is used to
describe the contribution of the Fe sublattice, the second term is
the additional Curie--Weiss contribution of the Ce sublattice to the
magnetic field $B(T)=2\pi f_{\mu}(T)/\gamma_\mu$ at the muon site.
Here, $\tilde{C}$ describes the hyperfine coupling constant of the
Ce moments with the muon spin. A fit of this function to the $\mu$SR
frequency $f_{\mu}(T)$ obtained for CeFeAsO is shown in
Fig.~\ref{img.orderparameter} and the two contributions are
highlighted as described in the legend. This simple model describes
the data reasonably well for temperatures between 10~K and up to
$T^{Fe}_N=137$~K, yielding $f_0=25.7(5)$~MHz, $\alpha=2.4(4)$,
$\beta=0.24(1)$. The constants $\tilde{C}=2.5(5)$~K, and
$\theta=-3(0.5)$~K were obtained by restricting the fit range to the
most relevant temperature region between 10 K and 50 K. The
exponents $\alpha$ and $\beta$ are close to the results obtained for
the other systems without the Curie--Weiss contribution. The
enhanced $f_0$ could be due to a slightly different muon site
compared to other \textit{R}FeAsO compounds. The deviation from this
behavior below 10~K is attributed to the growth of antiferromagnetic
correlations in the vicinity of $T_N^{Ce}$.

\begin{figure*}[htb]

\includegraphics[width=17.9cm,angle=0,clip]{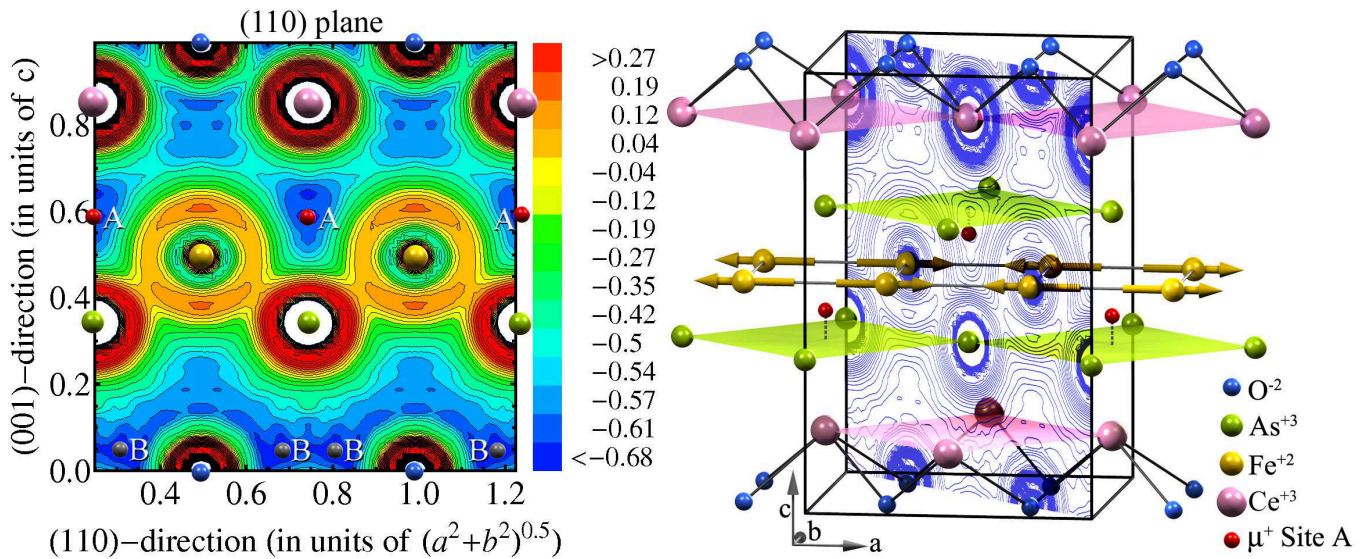}

\caption[]{Structure and electrostatic potential energy map of the
(110)--plane for a muon in the Cmma orthorhombic phase of CeFeAsO.
The A and B muon positions are marked by corresponding letters. The
potential energy is given in atomic units. The origin of the unit
cell has been moved for clarity.} \label{img.structure}

\end{figure*}

The exceptionally strong coupling of the Ce is already reasonable
considering the ground state properties of the \textit{R} ion in the
crystal electric field (CEF). The ground state of the Kramers ion
Ce$^{3+}$ is the J-multiplet $^2F_{5/2} $. The Sm$^{3+}$ Kramers ion
has $^6H_{5/2} $ as lowest $J$-multiplet with $J=5/2$. The
non--Kramers ion Pr$^{3+}$ stays in the $^3H_4 $ ground multiplet
with $J = 4$. The magnetic behavior of the \textit{R} ions can be
understood qualitatively from the susceptibility of free \textit{R}
ions:

\begin{equation}
\label{1.7} \chi_0 (T)=\frac{g_J^2 \mu_B^2 J(J+1)}{3T}
\end{equation}

\noindent Thus, because of the different g-factors ($g_J(\text{Ce})=6/7$, $g_J(\text{Sm})=2/7$, and
$g_J(\text{Pr})=4/5$) one expects at least one order of magnitude less induced Sm magnetic moment
compared to Ce (for equal magnitudes of 
Fe--Sm and Fe--Ce coupling constants, see section
\ref{subsec.Fe-Re-exchange}). Probably, due to this feature, we
observe very similar $\mu $SR spectra in the Sm, Pr, and La
compounds in the temperature range $T_N^{R}<T<T_N $. since the exact
sequence of CEF levels in PrFeAsO is not known up to now, it is more
difficult to predict its behavior.

The above qualitative discussion neglects higher CEF levels and the Ce--Ce interactions.
A more detailed and quantitative discussion of
the coupling of the Ce sublattice to the Fe sublattice using classical and quantum approaches will be
given in Sec.~\ref{sec.Fe-Re-interactions} and \ref{sec.magn-structures}.

\subsection{Rare earth magnetic order}

For \textit{R}~=~Pr and Ce the onset of the rare earth magnetic
order $11(1)$~K~and~4.4(3)~K, respectively causes a second order,
i.e. smooth decrease of the muon precession frequency $f_{\mu}(T)$
by approx. $\approx$2.2~MHz at $\approx$2~K in both cases, as can be
seen in Fig.~\ref{img.orderparameter}. However, in PrFeAsO the Pr
magnetic order is accompanied by a maximum in the width $\Delta B$
of the magnetic field distribution, i.e. the relaxation rate
$\lambda_T$ (see inset of Fig.~\ref{img.rate}) that rapidly
decreases at lower temperatures, whereas in CeFeAsO the Ce magnetic
order causes the field distribution to broaden monotonically. Note
that the magnetic field at the muon site caused by the static order
of the Ce sublattice leads to a decrease of the observed $\mu$SR
frequency while the Fe order induced magnetization of the Ce
sublattice above $T_N^{Ce}$ causes an increase of the frequency
$f_{\mu}(T)$. Therefore, it is evident that the Ce moments order in
a different structure than induced by the Fe magnetic system.

SmFeAsO shows first order Sm magnetic order, i.e. the Sm magnetic
order parameter is discontinuous at $T^{Sm}_N=4.66(5)$~K. Contrary
to the Ce and Pr magnetic order the Sm magnetic order causes the
appearance of two satellite muon precession frequencies
$f^{1,2}_\mu(T)=f^0_\mu(T)\pm8.0(5)$~MHz in addition to the main
frequency $f^0_\mu(T)$ observed above $T^{Sm}_N$. The two satellites
and the main frequency have signal fractions of
1.0(3):1.0(3):4.3(6). This shows that the Sm magnetic order has
different symmetry compared to the Fe magnetic order, i.e. it causes
a change of the magnetic unit cell. A detailed discussion of
\textit{R} magnetic structures and its interplay with the Fe
sublattice will be given in Sec.~\ref{sec.magn-structures},
\ref{sec.non-collinear-re-re} and \ref{sec.influence-Re-on-Fe}.

\section{Determination of the muon site}\label{sec.site}

To determine the contributions from Fe and \textit{R} magnetic order to the local magnetic field at the muon site it is
necessary to determine the muon site in the lattice. For this purpose we used a modified Thomas Fermi
approach\cite{Reznik08} and available structural data. The goal is to determine a self consistent distribution
of the valence electron density from which the electrostatic potential can be deduced. The local minima of this
potential at interstitial positions are regarded as possible stopping sites for muons. We verified the
applicability of our approach by comparison of numerical results with experimentally determined muon sites in
\textit{R}FeO$_{3}$\cite{Holzschuh83} and by a successful interpretation of $\mu$SR spectra of the complex magnetic
structures in layered cobaltites \textit{R}BaCo$_2$O$_{5.5}$.\cite{Luetkens-Co}

A potential map of CeFeAsO in the $Cmma$ orthorhombic phase was
calculated using structural data\cite{Zhao08a} at 1.4~K and is shown
in Fig.~\ref{img.structure}. The calculations have been done without
taking into account the host lattice relaxation around the muon. We
observe two types of possible muon positions which are labeled A and
B. The A type position is located on the line connecting the Ce and
As ions along the $c$--direction. Note that a similar location of
the point with deepest potential was calculated for LaOeAs in the
tetragonal phase using the general potential linearized augmented
plane-wave method and local density approximation.\cite{Takenaka78}
The A type position with coordinates (0,1/4,z$_A$) has 4g local
point symmetry mm2, i.e. the same as the \textit{R} sites. The B
type position has a general 16o local point symmetry with
(x,y,z$_B$) coordinates. For CeFeAsO the coordinates of A are
(0,1/4,0.41) and (0.44,0.19,0.04) for B. Note that the points B are
located at oxygen ions which are typical points for muons in many
oxides. The z$_{A}$ and z$_{B}$ coordinates vary in the third
decimal if Ce is replaced by La, Pr or Sm.

A second important result is the comparison of the calculated muon
precession frequency, i.e. the local magnetic dipole field at the
muon site with the experimentally determined muon precession
frequency $f_0=f_{\mu}(T\to 0)$ for LaFeAsO. With the experimentally
determined Fe magnetic moment \cite{Cruz08} of 0.36~$\mu_{\text{B}}$
we calculate {f}$_{0,A}=12.4$~MHz for site A and $f_{0,B}=1.3$~MHz
for site B. As shown by Klauss et al. in Ref.~\onlinecite{Klauss08}
and Section~\ref{sec.muon} of this work, in LaFeAsO two muon
frequencies are observed. One frequency with 23~MHz originating from
the mayor volume fraction ($P=0.7$) and one lower frequency with
3~MHz which develops from a strongly damped signal below
approximately 70~K. In \textit{R}FeAsO with \textit{R} = Sm, Pr, and
Ce only the high frequency is present in the $\mu$SR spectra. We
conclude that site A is the main muon site since this gives the
correct order of magnitude for $f_{0}$. Probably the site B is also
partially populated at low temperatures in the LaFeAsO. The fact
that we obtain a 46~{\%} smaller value than the experimental result
is reasonable since our calculation only takes into account local
dipole fields and neglects contact hyperfine contributions. Similar
differences are found in La$_{2}$CuO$_{4}$, the antiferromagnetic
parent compound of the 214--cuprate superconductors.\cite{Klauss04}
As will be shown below, the magnetic field caused by the Fe magnetic
order is directed along the crystallographic $c$--axis. This is in
agreement with recent $\mu$SR experiments on Fe pnictide single
crystals.\cite{luetkensInprep} In the following we use only muon
site A and a renormalization factor of 1.86 for the local field
caused by Fe magnetic ordering at this muon site to account for the
contact hyperfine field for all \textit{R}. This assumption is
justified from our calculation of the electronic charge density
distribution that shows very similar results for all \textit{R}.
This manifests itself in the almost identical position of muon site
A in compounds with different \textit{R}. In its turn this means
that the renormalization due to the contact hyperfine interaction,
which depends only on the electron density at the muon site, should
be nearly the same within this series.

\section{Symmetry analysis of Fe and \textit{R} magnetic order parameters in \textit{R}FeAsO}\label{sec.Fe-Re-interactions}

\subsection{Translational symmetry and magnetic modes in \textit{R}FeAsO compounds}\label{subsec.Fe-Re-symm}

Iron magnetic order in \textit{R}FeAsO sets in about 10--20~K below
the tetragonal to orthorhombic structural phase transition.
\cite{Cruz08} The space group of the paramagnetic phase is $Cmma$
with Fe and \textit{R} ions occupying 4b and and 4g position
respectively. Neutron studies of \textit{R}FeAsO compounds revealed
numerous magnetic Bragg peaks all of which can be reduced just to
three types of magnetic propagation
vectors.\cite{Cruz08,Kimber08,Zhao08a,Zhao08b,Qiu08,Huang08,Lynn08}
These vectors in an orthorhombic setting are:
\begin{eqnarray}
\mathbf K_{I}&=&(1,0,1/2)\nonumber\\
\mathbf K_{II}&=&(1,0,0)\label{1.1}\\
\mathbf K_{III}&=&(0,0,1/2).\nonumber
\end{eqnarray}


\noindent Here we use an orthorhombic primitive cell with the unit cell vectors $\mathbf a_O=2\tau_x
\mathbf e_a$, $\mathbf b_O=2\tau_y \mathbf e_b$, and $\mathbf c_O=2\tau_z \mathbf e_c$ where $\tau_x$,
$\tau_y$, and $2\tau_z$ are the distances between nearest neighbor Fe ions along a, b and c directions, denoted
by the unit vectors $\mathbf e_i$, respectively. Note that in this setting nuclear Bragg peaks of the $Cmma$ spacegroup are either $(2H,2K,L)$ or $(2H+1,2K+1,L)$. 
Also, a 'magnetic setting' with $\mathbf a_m=\mathbf a_O$,
$\mathbf b_m=\mathbf b_O$, and $\mathbf c_m=2\mathbf c_O$ is often used.

\begin{figure}[htb]
\includegraphics[width=\columnwidth]{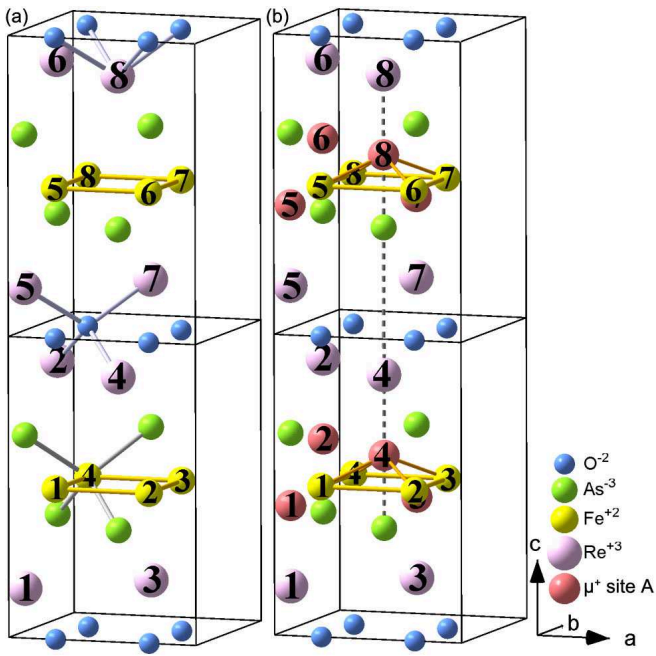}
\caption{Enumeration of positions used to describe different types of Fe and \textit{R} magnetic order. (a) \textit{R} and Fe
positions and b) muon site A.\label{img.num}}
\end{figure}

The primitive cells of the magnetic structures induced by each of the $\mathbf K_{\gamma}$ are not orthorhombic.
Each of the magnetic primitive cells contains four magnetically inequivalent \textit{R}--, Fe--, and muon--sites. To
analyze the symmetry of the magnetic structures in a unified way we will label these four positions with numbers
$\alpha = 1,2,3,4$, with corresponding ordered magnetic moments $\mathbf{m}_\alpha$ for the \textit{R}-- and
Fe--positions. To identify the different positions $\alpha$ in the magnetic cell with positions $\beta =
1, 2,\ldots, 8$ in the nuclear cell (doubled along the $c$--direction) we use the mapping $\{\beta_1, \beta_2\}
\mapsto \alpha$ where $\alpha$ is one of the four possible positions in the magnetic primitive cell and
$\beta_i$ a position in the nuclear cell as shown in Fig.~\ref{img.num}.

The $\mathbf K_I$ propagation vector induces a magnetic lattice of $I_{c}mmm$ type in which the primitive
cell is built on the translations $\mathbf a_1=\mathbf a_O$, $\mathbf a_2=\mathbf b_O$, $\mathbf a_3=1/2(\mathbf
a_O+\mathbf b_O)+\mathbf c_O$ and the anti--translation along the z--axis ${\mathbf a}'=\mathbf c_O$.\cite{X1}
Thus, in $\mathbf K_I$--type magnetic structures the following positions are equivalent and mapped as denoted by
'$\mapsto$': $\{1,7\}\mapsto1$, $\{2,8\}\mapsto2$, $\{3,5\}\mapsto3$, and $\{4,6\}\mapsto4$.

The primitive cell of the $P_{C}mmm$ magnetic lattice induced by the $\mathbf K_{II}$ propagation vector is
constructed from the translations $\mathbf a_1=\mathbf a_O$, $\mathbf a_2 =\mathbf b_O$, $\mathbf a_3 =\mathbf
c_O $ and the anti translation ${\mathbf a}'=1/2(\mathbf a_O + \mathbf b_O )$.\cite{X1} Then we obtain the
identity of the positions: $\{1,5\}\mapsto1$, $\{2,6\}\mapsto 2$, $\{3,7\}\mapsto3$, and $\{4,8\}\mapsto4$.

Finally, for the $\mathbf K_{III}$ propagation vector the magnetic lattice is $C_{c}mmm$ and the magnetic
primitive cell can be built from the translations: $\mathbf a_1 =1/2(\mathbf a_O +\mathbf b_O )$, $\mathbf a_2
=1/2(\mathbf a_O-\mathbf b_O )$, $\mathbf a_3 =2\mathbf c_O $.\cite{X1} Therefore we have the mappings:
$\{1,3\}\mapsto1$, $\{2,4\}\mapsto2$, $\{5,7\}\mapsto3$, and $\{ 6,8\}\mapsto4$.

In order to investigate the magnetic interactions in different \textit{R} and Fe magnetic structures with the
translational symmetry of the propagation vector $\mathbf K_\gamma$ ($\gamma \in \{0, I, II, III\}$) it is
necessary to calculate the magnetic modes, i.e. basis functions of the irreducible representations (IR) of the
propagation vector little groups $G_\gamma $. These groups are the same for all propagation vectors (\ref{1.1}) and they are
also identical with the groups for ferromagnetic order, i.e. propagation vector $\mathbf K_0 =(0,0,0)$.\cite{X2}
Following a method described by Bertout, and Izyumov and Naish we introduce possible magnetic modes $\mathbf{F}$
and $\mathbf{L}_i$, $i$=1,2,3 as linear combinations of Fe (or \textit{R}) sublattice magnetic moments $\mathbf
{m}_\alpha $, where $\alpha $ denotes a particular atom (see above) in the respective magnetic primitive cell:\cite{X3,X4}

\begin{eqnarray}
\mathbf {F}&=&
1/4(\mathbf {m}_1 +\mathbf {m}_2 +\mathbf {m}_3 +\mathbf {m}_4)\nonumber\\
\mathbf {L}_1 &=&1/4(\mathbf {m}_1 +\mathbf {m}_2 -\mathbf {m}_3 -\mathbf {m}_4 )\nonumber\\
\mathbf {L}_2 &=&1/4(\mathbf {m}_1 -\mathbf {m}_2 +\mathbf {m}_3 -\mathbf {m}_4 )\nonumber\\
\mathbf {L}_3 &=&1/4(\mathbf {m}_1 -\mathbf {m}_2 -\mathbf {m}_3 +\mathbf {m}_4).\label{1.2}
\end{eqnarray}

\noindent The components of these linear combinations are the basis functions of the eight one-dimensional IR
$\tau_\nu$ with $1\leq\nu\leq8$ for the $G_0$ group (see table 32 in Ref.~\onlinecite{X2}). The results of the
symmetry analysis for the magnetic moments located at the 4b positions (Fe ions) and 4g positions (\textit{R} ions and
muon sites) are shown in Table~\ref{tab.X1}.

\begingroup
\squeezetable
\begin{table*}[htb]
\caption{Magnetic modes of the magnetic propagation vectors $\mathbf K_\gamma$, i.e. basis functions of the irreducible representations $\tau_\nu$ (see
text). The indexes indicate the non--zero Cartesian components of the vectors (\ref{1.2}), i.e. for example $L_{1z}$ is the $z$--component of $\mathbf L_1$.\label{tab.X1}}
\begin{ruledtabular}
\begin{tabular}{cdddddddd|cc|cc|cc|cc}
&\multicolumn{8}{c|}{\begin{minipage}[c]{80mm} \centering{$Cmma$ symmetry elements \newline \footnotesize{(index 1 denotes an improper translation along the y--axis)}} \end{minipage}}&
\multicolumn{2}{c|}{\begin{minipage}[c]{12mm} \centering{$\mathbf K_0$ (0,0,0)} \end{minipage}} &\multicolumn{2}{c|}{\begin{minipage}[c]{23mm} \centering{$\mathbf K_I$ ($\pi/\tau _x$,0,$\pi/2\tau _z$)} \end{minipage}} &\multicolumn{2}{c|}{\begin{minipage}[c]{15mm} \centering{$\mathbf K_{II}$ ($\pi/\tau _x$,0,0)} \end{minipage}} &\multicolumn{2}{c}{\begin{minipage}[c]{19mm} \centering{$\mathbf K_{III}$ (0,0,$\pi/2\tau _z$)} \end{minipage}} \\
\hline
IR&1&\multicolumn{1}{c}{2$_x$}&\multicolumn{1}{c}{2$_{1y}$}&\multicolumn{1}{c}{2$_{1z}$}&\multicolumn{1}{c}{$\bar{1}$}&\multicolumn{1}{c}{$m_x$}&\multicolumn{1}{c}{$m_{1y}$}&\multicolumn{1}{c|}{$m_{1z}$}&\textbf{Fe}&\textbf{\textit{R}} \& $\mu^+$&\textbf{Fe}&\textbf{\textit{R}} \& $\mu^+$&\textbf{Fe}&\textbf{\textit{R}} \& $\mu^+$&\textbf{Fe}&\textbf{\textit{R}} \& $\mu^+$\\
\hline
$\tau_1$&1&1&1&1&1&1&1&1&--&--&$L_{3y}$&--&$L_{1x}$&--&$L_{3z}$&-- \\
$\tau_2$&1&1&1&1&-1&-1&-1&-1&--&$L_{2z}^R$&$L_{1y}$&$L_{1z}^R$&${L_{3x}}$&${L_{3z}^R}$&$L_{1z}$&$L_{1z}^R$ \\
$\tau_3$&1&1&-1&-1&1&1&-1&-1&$F_x$&$F_x^R$&$L_{3z}$&$L_{3x}^R$&--&$L_{1x}^R$&$L_{3y}$&$L_{3x}^R$ \\
$\tau_4$&1&1&-1&-1&-1&-1&1&1&$L_{2x}$&$L_{2y}^R$&$L_{1z}$&$L_{1y}^R$&--&$L_{3y}^R$&$L_{1y}$&$L_{1y}^R$ \\
$\tau_5$&1&-1&1&-1&1&-1&1&-1&$F_y$&$F_y^R$&--&$L_{3y}^R$&$L_{1z}$&$L_{1y}^R$&$L_{3x}$&$L_{3y}^R$ \\
$\tau_6$&1&-1&1&-1&-1&1&-1&1&$L_{2y}$&$L_{2x}^R$&--&$L_{1x}^R$&$L_{3z}$&$L_{3x}^R$&$L_{1x}$&$L_{1x}^R$ \\
$\tau_7$&1&-1&-1&1&1&-1&-1&1&$F_z$&$F_z^R$&${L_{3x}}$&${L_{3z}^R}$&$L_{1y}$&$L_{1z}^R$&--&$L_{3z}^R$ \\
$\tau_8$&1&-1&-1&1&-1&1&1&-1&$L_{2z}$&--&$L_{1x}$&--&$L_{3y}$&--&--&--\\
\end{tabular}\end{ruledtabular}
\end{table*}
\endgroup

In accordance with neutron data the iron magnetic subsystem orders antiferromagnetically along the $a$--direction
and ferromagnetically along the $b$--direction with doubling of the unit cell along the $c$--axis for the \textit{R} = La
compound (magnetic propagation vector $\mathbf K_I$) and without doubling for \textit{R} = Ce and Pr (magnetic propagation vector $\mathbf
K_{II}$).\cite{Lynn08} In all these structures Fe magnetic moments are directed along the $a$--axis. These
magnetic structures are described by $L_{3x}(\mathbf K_I)$ and $L_{3x}(\mathbf K_{II})$ non--zero magnetic order
parameters shown in Table \ref{tab.X1}.

The onset of a given type of magnetic order lowers the symmetry of the paramagnetic phase and creates a magnetic
symmetry. The orientation of the magnetic moments and the distribution of local magnetic fields in the magnetic
cell have the same symmetry. To determine the orientation of the magnetic field at a particular muon site
we assign an artificial magnetic degree of freedom to this site. The set of magnetic degrees of freedom for
different points (Wyckoff positions) forms a magnetic representation. A standard decomposition of this
magnetic representation into irreducible representations of the space group allows the analysis of the symmetry
of magnetic field distributions at the muon sites. This symmetry must be compatible with the given space group
and the same as the symmetry of the magnetic order parameter. In other words, the symmetry of the magnetic field distribution at the muon site must belong to the same IR as
the magnetic order parameter.

From Table~\ref{tab.X1} it becomes self--evident that the $L_{3x}(\mathbf K_{II})$--type of magnetic structure of
iron moments creates $L^R_{3z}(\mathbf K_{II})$--type molecular fields at the rare earth and A--type muon sites.
Therefore, a magnetic field distribution of any symmetry that induces a staggered-type ordering of the \textit{R}
magnetic moments leads to a magnetic field distribution at the A--type muon site of same symmetry. Due to
symmetry reasons, some types of Fe magnetic order, for instance $L_{3y}(\mathbf K_I)$, do not create magnetic
fields at the 4g positions. Thus, the presence of $\mu$SR signals in the Fe antiferromagnetically (AFM) ordered phase of \textit{R} = La, Pr, Ce and Sm
compounds excludes the possibility for AFM magnetic order along the $a$--axis and ferromagnetic (FM) order along $b$--axis with iron magnetic
moments directed along $b$--axis.

\begin{figure}[htb]
\includegraphics[width=\columnwidth]{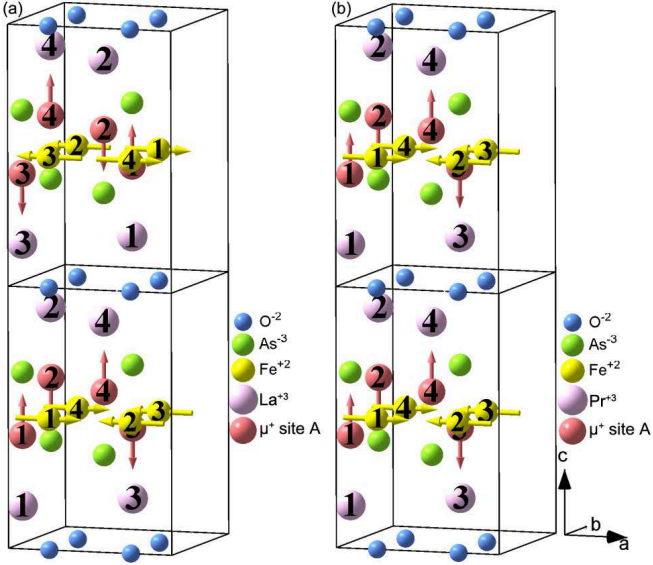}
\caption{(a) L$_{3x}$(K$_I$)--type Fe magnetic order realized in
LaFeAsO and probably in SmFeAsO and its magnetic dipole field
direction at the muon site A. (b) L$_{3x}$(K$_{II}$) order observed
in CeFeAsO and
PrFeAsO.\label{img.Fig-KI-and-KII-Fields-on-muon-above}}
\end{figure}

The $L_{3x}(\mathbf K_I)$--type of Fe order and the corresponding direction of the magnetic field at the A--type
muon site are shown in Fig.~\ref{img.Fig-KI-and-KII-Fields-on-muon-above}. The dipole fields at the A--type muon
position for all types of Fe and \textit{R} magnetic order are given in the Appendix A.

\subsection{Symmetry of the Fe--\textit{R} magnetic interactions}\label{subsec.Fe-Re-exchange}

To start with the strongest exchange interaction we analyze the symmetry of permutation modes (\ref{1.1})
or exchange multiplets for Fe and \textit{R} magnetic subsystems.\cite{X3} The permutation modes for the different
translational symmetries $\mathbf K_\gamma$ are listed together with the corresponding irreducible
representations in table~\ref{tab.X2}.

\begingroup
\squeezetable
\begin{table}[htb]
\caption{Permutation modes, i.e. exchange multiplets of the four magnetic propagation vectors.\label{tab.X2}}
\begin{ruledtabular}
\begin{tabular}{ccc|cc|cc|cc}
&\multicolumn{2}{c|}{\begin{minipage}[c]{12mm} \centering{$\mathbf K_0$ (0,0,0)} \end{minipage}} &\multicolumn{2}{c|}{\begin{minipage}[c]{21mm} \centering{$\mathbf K_I$ ($\pi/\tau _x$,0,$\pi/2\tau _z$)} \end{minipage}} &\multicolumn{2}{c|}{\begin{minipage}[c]{15mm} \centering{$\mathbf K_{II}$ ($\pi/\tau _x$,0,0)} \end{minipage}} &\multicolumn{2}{c}{\begin{minipage}[c]{19mm} \centering{$\mathbf K_{III}$ (0,0,$\pi/2\tau _z$)} \end{minipage}} \\
\hline
IR&\textbf{Fe}&\textbf{\textit{R}} &\textbf{Fe}&\textbf{\textit{R}} &\textbf{Fe}&\textbf{\textit{R}} &\textbf{Fe}&\textbf{\textit{R}} \\
\hline
$\tau _1 $ &$\mathbf {F}$&$\mathbf {F}^R $&--&$\mathbf {L}_1^R $&--&$\mathbf {L}_1^R $&--&$\mathbf {L}_3^R $ \\
$\tau _2 $&$\mathbf {L}_2$&--&--&--&--&--&--&-- \\
$\tau _3 $&--&--&--&--&$\mathbf {L}_1$&--&--&-- \\
$\tau _4 $&--&--&--&--&$\mathbf {L}_3$&--&--&-- \\
$\tau _5 $&--&--&$\mathbf {L}_3$&--&--&--&$\mathbf {L}_3$&-- \\
$\tau _6 $&--&--&$\mathbf {L}_1$&--&--&--&--&-- \\
$\tau _7 $&--&--&--&--&--&--&--&-- \\
$\tau _8 $&--&$\mathbf {L}_2^R $&--&$\mathbf {L}_3^R $&--&$\mathbf {L}_3^R $&$\mathbf {L}_1$&$\mathbf {L}_1^R $\\
\end{tabular}\end{ruledtabular}
\end{table}
\endgroup

Since there are no permutation modes of the Fe-- and \textit{R}--subsystems which belong to the same irreducible
representation there are no Heisenberg exchange interactions between the \textit{R}-- and Fe--subsystems for the
case of $\mathbf K_I$ or $\mathbf K_{II}$ translational symmetry of the Fe magnetic order. This interaction exists
only for magnetic structures with $\mathbf K_0$ and $\mathbf K_{III}$ translational symmetry. The respective Fe--\textit{R} magnetic exchange Hamiltonian has the form:

\begin{eqnarray}
\mathcal H_{ex}^{Fe-R}&=&\ldots+J_0^{Fe-R}(\mathbf K_0)(\mathbf{F} \cdot \mathbf{F}^R) \nonumber\\
& &+ J_1^{Fe-R}(\mathbf K_{III})(\mathbf{L}_1(\mathbf K_{III})\mathbf{L}^R_1(\mathbf K_{III})) \label{1.3}
\end{eqnarray}

\noindent However, for the cases of $\mathbf K_I $ and $\mathbf K_{II} $ translational symmetry
the Fe and \textit{R} subsystems can interact by the non--Heisenberg exchange (see Table~\ref{tab.X2}). The part of these anisotropic Fe--\textit{R} interactions relevant for the following
considerations is given below:

\begin{eqnarray}
\mathcal H_{an-ex}^{Fe-R}&=&\ldots
+ I_{3xz}^{Fe-R} (\mathbf K_I )L_{3x} (\mathbf K_I)L_{3z}^R(\mathbf K_I ) + \nonumber\\
 & & + I_{3zx}^{Fe-R} (\mathbf K_I )L_{3z} (\mathbf K_I )L_{3x}^R (\mathbf K_I ) + \nonumber\\
 & & + I_{3xz}^{Fe-R} (\mathbf K_{II} )L_{3x} (\mathbf K_{II} )L_{3z}^R (\mathbf K_{II}) + \nonumber\\
 & & + I_{3zx}^{Fe-R}(\mathbf K_{II} )L_{3z} (\mathbf K_{II} )L_{3x}^R (\mathbf K_{II} ) + \nonumber\\
 & & + \ldots \label{1.4}
\end{eqnarray}

\begin{figure}[htb]
\includegraphics[width=\columnwidth]{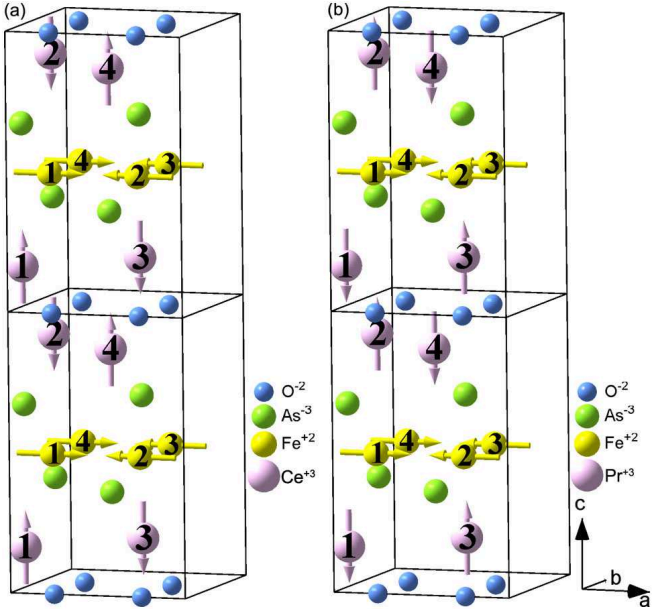}
\caption{Direction of the magnetic moment on the \textit{R} site
induced by the L$_{3x}$(K$_{II}$) Fe order in a) CeFeAsO and b)
PrFeAsO. The sign of the coupling constant $I_{3xz}^{Fe-R}$ is
different in the Ce and the Pr compound resulting in opposite
orientation of the induced moments.
\label{img.Fig-MagMoments-above}}
\end{figure}

\noindent Consequently, the onset of the $L_{3x}(\mathbf K_{I/II})$ type of Fe magnetic order creates an
effective staggered magnetic field at the \textit{R} site along the z--direction. The magnitude of this field is
proportional to the value of the Fe--\textit{R} coupling constant $I_{3xz}^{Fe-R} $ and the iron subsystem order
parameter $L_{3x} $ determines its temperature dependence. In the following we will show that this field exceeds
a respective dipole field at the \textit{R} sites by at least one order of magnitude. The mutual orientation of the
$L_{3x}$ and $L_{3z}^R$ vectors depends on the sign of the Fe--\textit{R} coupling constant $I_{3xz}^{Fe-R}$. The
orientation of the induced magnetic moment on the \textit{R} site is shown in Fig. \ref{img.Fig-MagMoments-above}.

The magnitude of the $z$--component of the rare earth magnetic moments $m_{\alpha z}^R \;(\alpha = 1 - 4)$ is determined by the exchange field
and the \textit{R} ion crystal electric field (CEF).

\section{Determination of the magnetic structures and the Fe--\textit{R} coupling constants from the temperature dependence of the $\mu$SR response}\label{sec.magn-structures}

Below the N\'eel temperature the $\mu$SR response of \textit{R}FeAsO
is mainly determined by magnetic dipole fields created by the Fe and
\textit{R} subsystems on the A-type muon site. It can be calculated
by:
\begin{eqnarray}
B_z(A)& = & 2.58 \cdot L_{3x}(\mathbf K_I ) + 1.45 \cdot L^R_{3z}(\mathbf K_I) \nonumber\\
B_z(A) & = & 2.57 \cdot L_{3x}(\mathbf K_{II}) + 1.48 \cdot L^R_{3z}(\mathbf K_{II}). \label{1.6}
\end{eqnarray}
\noindent If the $L_{ij}(\mathbf K_{\gamma})$ are given in units of
\mb{} (see equation (\ref{1.2})) the magnetic dipole field $B_z(A)$
is obtained in units of 0.1~T. These equations demonstrate a
surprisingly close agreement of dipole fields for Fe magnetic order
with ($\mathbf K_I$) and without ($\mathbf K_{II}$) doubling along
z--axis. Note that the constants in equations (\ref{1.6}) only vary
in the third decimal if Ce is replaced by La, Pr or Sm. This
explains the similar $\mu$SR spectra of LaFeAsO with $L_{3x}
(\mathbf K_I )$--, and CeFeAsO and PrFeAsO both with $L_{3x}(\mathbf
K_{II})$--type of Fe antiferromagnetic order. The high temperature
fit of the $\mu$SR frequency gives roughly the same saturation value
$f_0=\gamma_\mu / 2\pi B = 23.0(5)$~MHz. This corresponds to a value
of 0.36~\mb{} for the Fe saturation moment if one takes into account
contact hyperfine fields by applying the renormalization factor of
1.86 as in LaFeAsO. Because of the similar order parameters of the
Fe magnetic order found in SmFeAsO, LaFeAsO, and PrFeAsO we conclude
that a $L_{3x}$--type of Fe antiferromagnetic order is also realized
in SmFeAsO. However, $\mu$SR studies alone cannot distinguish
between the possible translational symmetries, i.e. $\mathbf K_I$
and $\mathbf K_{II}$ of the order parameter.

\subsubsection{CeFeAsO}

\begin{figure*}[htb]
\includegraphics[width=11cm]{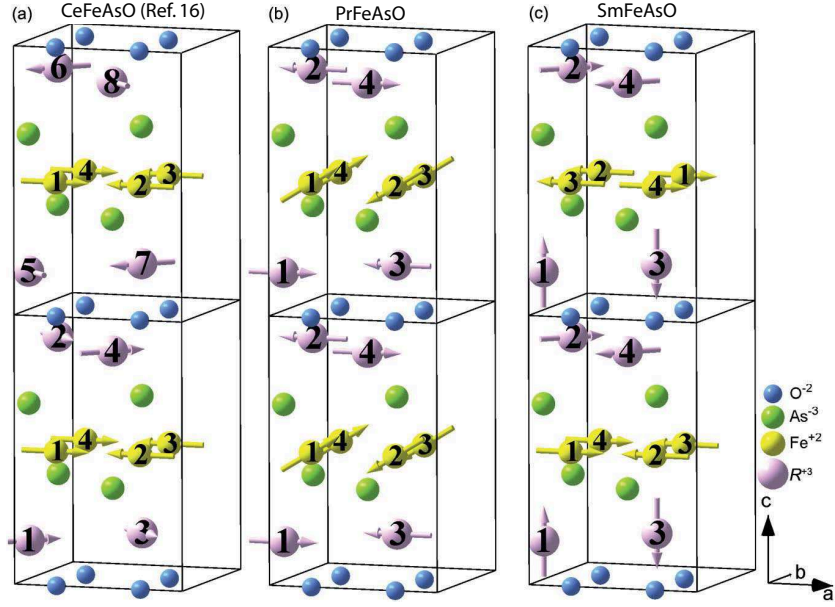}
\caption{(a) Collinear Fe magnetic order ($\mathbf K_{II}$) and
non--collinear Ce order in CeFeAsO as proposed by Zhao et al.
\cite{Zhao08a} which is in accordance with the $\mu$SR measurements.
(b) Magnetic structure of the Pr and Fe subsystems with $\mathbf
K_{II}$ translation symmetry which satisfactorily describes $\mu$SR
and Mössbauer data. Non--zero are the $L_{3x} \cos \theta + L_{3z}
\sin \theta $ of Fe and $L_{3x}^R$ of Pr order parameters. The Fe
and Pr subsystems remain in the $L_3$ and $L_{3}^R$ type of exchange
magnetic order, respectively. (c) The proposed Sm structure which
explains the observed $\mu$SR spectra. Fe magnetic ordering has
$\mathbf K_I$ translational symmetry with a doubling along the
$c$--axis, while the Sm magnetic order possesses $\mathbf K_{II}$
translational symmetry without a doubling along the
$c$--axis).\label{img.Fig-MagMoments-below}}
\end{figure*}

The $\mu$SR response drastically changes below the Ce ordering
temperature $T_N^{Ce}$. Contrary to the naive expectation that the
Ce magnetic order would increase the local magnetic field at the
muon site and therefore the $\mu$SR frequency our experimental
results show a decrease of the $\mu$SR frequency below $T_N^{Ce}$.
According to our analysis of the magnetic field distribution at the
A--type muon site this behavior indicates a breaking of the
$L_{3z}^R$ symmetry of the induced Ce order. Neutron diffraction
studies of CeFeAsO revealed the appearance of $\mathbf
K_{III}$--type magnetic Bragg peaks below $T_N^{Ce}$ in addition to
$\mathbf K_{II}$--type peaks also present above
$T_N^{Ce}$.\cite{Zhao08a} The Ce magnetic structure proposed by Zhao
et al. is a non--collinear arrangement of Ce moments in the
$ab$-plane.\cite{Zhao08a} This structure is shown in
Fig.\ref{img.Fig-MagMoments-below} a) and can be described as a
linear combination of two order parameters: $L_{1x}^R (\mathbf
K_{III} )+L_{3y}^R (\mathbf K_{II})$. This new type of \textit{R}
order produces the following magnetic field at the A-type muon site
for temperatures below $T_N^{Ce}$:

\begin{eqnarray}
B_x (A)&=&-0.40 \cdot L^{Ce}_{1x}(\mathbf K_{III})\nonumber\\
B_y (A)&=&-0.88 \cdot L^{Ce}_{3y}(\mathbf K_{II})\nonumber\\
B_z (A)&=& 2.57 \cdot \frac{f_0 }{13.55} + 1.48 \cdot L^{Ce}_{3z}(\mathbf K_{II}) \label{1.8}
\end{eqnarray}

Here $f_0$=25.7~MHz accounts for the contribution of the iron subsystem, which, we suppose, preserves its magnetic structure found above $T_N^{Ce}$. Despite different translational symmetry of the Ce order parameters, the magnetic fields at all eight muon sites
have the same modulus (\ref{1.8}), therefore no additional frequencies appear at temperatures below $T_N^{Ce}$.
With the Ce magnetic moment of 0.83~\mb{} at 1.7~K reported by Zhao et al. \cite{Zhao08a} and supposing
$L_{3z}^{Ce}$=0 we then obtain a $\mu$SR frequency of $f=26.9$~MHz instead of the experimentally observed 29.8 MHz
at 1.9~K. Also the experiment reveals a remarkable decrease of the observed $\mu$SR frequency compared to maximal
value of 32~MHz at 4~K. The decrease of the $\mu$SR frequency is not as rapid as expected, it develops only gradually below $T_N^{Ce}$. As follows from equation (\ref{1.8}), the magnetic dipole fields at the muon sites
created by the Ce magnetic moments lie in the $ab$--plane and are rather weak compared to the dipole field created by the Fe
moments. The slow decrease of the $\mu$SR frequency below $T_N^{Ce}$ therefore reflects the gradual
disappearance of the Fe induced $L_{3z}^{Ce}$ type magnetization of the Ce moments in the AFM Ce phase below $T_N^{Ce}$. One
can conclude that the Ce--Ce coupling constant which creates the $L^{Ce}_{1x} (\mathbf K_{III} )+L^{Ce}_{3y} (\mathbf
K_{II} )$ AFM Ce order parameter cannot immediately suppress the induced $L_{3z}^{Ce}$ magnetization. As will be shown
below the Ce--Ce coupling constant $J_{3z}^{Ce} (\mathbf K_{II})$ is of the same order as the Fe--Ce coupling constant $I_{3xz}^{Fe-Ce} (\mathbf K_{II})$. Using
equation (\ref{1.8}) we estimate the remanent order parameter $L_{3z}^{Ce}$=0.14$\mu_B$ at 1.9~K.

\begin{figure}[htb]
\includegraphics[width=8.6cm]{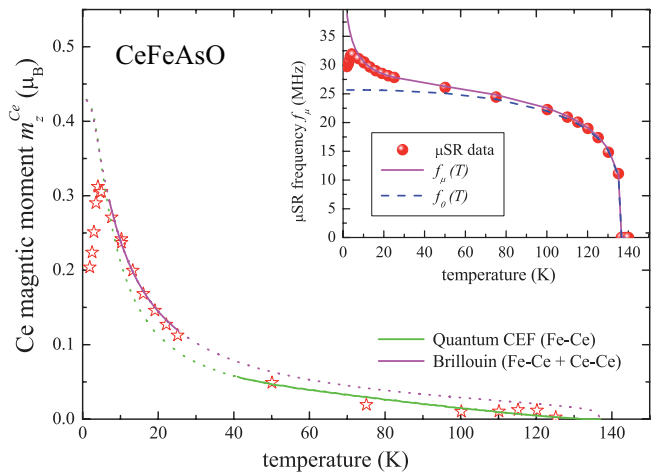}
\caption{Temperature dependence of the induced Ce magnetic moment $m_z^{Ce}(T)$ for $T>T^{Ce}_N$.
$m_z^{Ce}(T)$ has been extracted from the $\mu$SR frequency by subtracting the contribution from the Fe
subsystem (inset). The data are fitted by a quantum and a Brillouin  approach valid in different temperature
regions, see text.\label{MCe}}
\end{figure}

Due to the significant contribution of Ce dipole fields, the $\mu$SR
frequency $f_\mu(T)$ is unusually high already far above $T_N^{Ce}$
in CeFeAsO. This has been discussed phenomenologically in
Sec.~\ref{sec.results}. Here we carry out a more elaborate analysis
of this observation. The lowest crystal electric field (CEF)
excitations of CeFeAsO and SmFeAsO consist of three Kramers
doublets. With the onset of antiferromagnetic iron order these
doublets split. Inelastic neutron scattering studies of CeFeAsO
revealed that in the paramagnetic phase the energies of the first
and the second excited doublets are 216~K and 785~K
respectively.\cite{Chi08} The splitting at $T=7$~K amounts to 10.8~K
for the ground state doublet and 37.1~K and 66.1~K for the first
excited levels. Using these experimentally deduced parameters and
the wave functions of the corresponding CEF levels allows us to
determine the Fe--Ce and Ce--Ce exchange interaction coupling
constants. In the Fe antiferromagnetic phase above $T_N^{Ce}$, they
are obtained by fitting the theoretical temperature dependence of
the Ce magnetic moment $m_z^{Ce} (T)$ to the experimental data. In
Fig.~\ref{MCe} the experimental value of the Ce single ion magnetic
moment $m_z^{Ce} (T)$ is shown. It is calculated from the $\mu$SR
frequency $f_{exp}$ with the help of equation (\ref{1.6}) after
subtracting the Fe contribution $f_0(T)$ (see inset of
Fig.~\ref{MCe}):
\begin{equation}
L_{3z}^{Ce}(T)= m_z^{Ce}(T)=[f_{exp}(T)-f_0(T)]/13.55\cdot 1.48 \label{mCe}
\end{equation}
In this effective field approximation the Fe--Ce coupling constant $I_{3xz}^{Fe-Ce}(K_{II})$ and the Ce--Ce coupling constant $J_{3z}^{Ce} (K_{II} )$ both create a staggered effective field $B_z^{eff}(T)$ on the Ce sites.
Its interaction with the Ce magnetic moments lifts the degeneracy of the Kramers doublets. The corresponding
effective field on the Ce site is given by:
\begin{equation}\label{Hefm}
B_z^{eff}(T) = B_z^{Fe-Ce} (T) + J_{3z} m_z^{Ce} (T)
\end{equation}
where $B_z^{Fe-Ce}(T)=I_{3xz}^{Fe-Ce}(K_{II})L_{3x}(T)$ and $J_{3z}=2J_{3z}^{Ce}(K_{II})$
with $I$ and $J$ in units of T/$\mu_B$. We neglect contributions from Fe--Ce and Ce--Ce dipole fields to the
splitting of Kramers doublets under the assumption that the Fe--Ce and Ce--Ce (non--)Heisenberg exchange fields (see equation  \ref{Hefm}) are much stronger. The Fe--Ce staggered field $B_z^{Fe-Ce}$ has to have the same temperature dependence as $f_0 (T)$, i.e. the
averaged value of the iron order parameter $L_{3x}(T)$. Therefore, it can be modeled as
$B_z^{Fe-Ce}(T)=B_0\cdot[1-(T/T_c)^\alpha]^\beta$ with $\alpha=2.4$ and $\beta=0.24$ as determined in
Sec.~\ref{sec.results}.

According to the analysis of inelastic neutron scattering data \cite{Chi08} the lowest Kramers doublet possesses the
orbital momentum $J_z=\pm 1/2$ and does not contain any admixtures of the $J_z=\pm3/2, \pm 5/2$ states. One can
estimate the magnitude of the effective field from the splitting of the ground Kramers doublet at $T=7$~K by
using the relation:
\begin{equation}\label{delta}
g_J \mu _BB_z^{eff}(T)=k_B\Delta
\end{equation}
where $g_J=6/7$ is the g-factor of the free $Ce^{3+}$ ion and $\Delta=10.8$~K the splitting of the ground state
doublet. If we account for just the Fe--Ce interaction it yields $B_0=18.75$~T. This large Fe--Ce effective field
indicates that dipole--dipole interactions (see Appendix B) do not play a significant role for the magnetization
of the Ce sublattice by the ordered Fe sublattice above $T^{Ce}_N$.

The temperature dependence of the Ce moment in the effective field has been calculated in a quantum mechanical approach
according to equation (10) in Ref.~\onlinecite{Mesot97} by calculating the thermal population of all
experimentally determined CEF levels.\cite{Chi08} The result is shown in Fig.~\ref{MCe} as a green line. From
our calculation it is clear that there is a deviation from Curie-Weiss behavior due to a sizable contribution to
$m_z^{Ce}$ from higher Kramers doublets for temperatures above 50~K. In Sec.~\ref{sec.results} we did a
Curie-Weiss approximation of $m_z^{Ce}$ by using Eq.~(\ref{eqn.ce}). A Curie-Weiss description
neglects higher crystal field levels. This is the reason why we restricted the fit to the temperature range
between 10 and 50~K in the first approximation done in Sec.~\ref{sec.results}.

On the other hand, a systematic deviation between $m_z^{Ce}$ obtained from the quantum approach and the
experimental data is observed in the low temperature region between 10 and 50~K. This indicates that the Ce--Ce
interaction, which is neglected by this approach is significant in this temperature region and has to be taken
into account for a proper description of the data. Therefore, we fit the Ce magnetic moment with contributions
from the Fe--Ce and Ce--Ce exchange interactions by using the usual Brillouin function. In the case of S=1/2 this
leads to the following equation:

\begin{equation}\label{Brill}
m_z^{Ce}(T)=\frac{g_J\mu _B }{2}\tanh\left({\frac{g_J \mu _B }{2k_B T}}\,B_z^{eff}(T)\right)
\end{equation}

\noindent where $B_z^{eff}$ is given in Eq.~(\ref{Hefm}) which should fulfill the boundary condition given by Eq.~(\ref{delta}).
This approach neglects the higher crystal field levels and therefore can only be applied at temperatures below
50~K. From a fit in the low temperature region we obtain the microscopic parameters $B_0=26.8(5)$~T and
$J_{3z}=-24.3(5)$~T/$\mu_B$. The fit is shown as a magenta line in the Fig.~\ref{MCe}. Using 0.41~$\mu_B$ for
iron saturated moment one can estimate a rather large Fe--Ce coupling constant $I_{3xz}^{Fe-Ce}(K_{II})=65.3$~T/$\mu_B$. The negative sign of the Ce--Ce coupling constant $J_{3z}^{Ce}(K_{II})=-12.15$~T/$\mu_B$ indicates
that the $L_{3}^{Ce}$ type of Ce exchange magnetic order is energy favorable. From $T_{N}^{\textit{R}}$=4.4 K one can
estimate the value of the $J_{3y}^{Ce}(K_{II})$ coupling constant. It is responsible for the appearance of the
in-plane $L_{3y}^{Ce}$ order parameter.\cite{Zhao08a} This rough estimate gives
$J_{3y}^{Ce}(K_{II})=-14.9$~T/$\mu_B$. Note, that the big difference between Ce--Ce exchange coupling constants
along z- and y--directions points to a strong easy-plane anisotropy in the Ce subsystem.

In conclusion, for CeFeAsO we found a sizable staggered
magnetization of the Ce ions induced by the Fe subsystem already far
above $T^{Ce}_N$ which amounts to approximately $0.3$~\mb{} near to
$T^{Ce}_N$. With the help of a symmetry analysis as well as
classical and quantum mechanical approaches we were able to deduce
Fe--Ce and Ce--Ce exchange coupling constants. We demonstrated that
in the Ce compound the Fe--Ce non-Heisenberg exchange interaction is
exceptionally strong and of the same order as the Ce--Ce exchange
interaction. The observed $\mu$SR frequency in the Ce ordered phase
is in agreement with the magnetic structure proposed by Zhao et
al.,\cite{Zhao08a} but a small component from the $L_{3z}^{Ce}$ type
Ce magnetic order induced by the Fe sublattice has to be included to
fully explain our data below $T_N^{Ce}$.

\subsubsection{PrFeAsO}

The ground state $^3H_4$ multiplet ($J = 4$) of the non--Kramers ion
Pr$^{3+}$ in the $C_{2v}$ low symmetry crystal field splits into
nine singlets. Since the exact structure of the CEF levels is not
known, the Pr$^{3+}$ magnetic susceptibility cannot be estimated.
The temperature dependence of the $\mu$SR frequency in PrFeAsO is
similar to the one observed in LaFeAsO. This suggests that the
induced Pr moment is not as strong as in CeFeAsO. However, the
comparison of the high temperature fit of the $\mu$SR frequency for
La, Ce, Sm, and Pr compounds yields the lowest value of
$f_{0}=22$~MHz for the Pr compound. Additionally, there is a small
but systematic deviation between $\mu$SR and Mössbauer data for
PrFeAsO in the Fe AFM phase. This demonstrates that also in PrFeAsO
a Fe induced ordered Pr moment is present in the Fe AFM phase. One
can conclude that, in contrast to the Ce compound, the Fe--Pr
coupling constant $I_{3xz}^{Fe-Pr}(\mathbf K_{II})$ is small and
negative. According to equation (\ref{1.6}), the fields created by
the Fe order parameter $L_{3x}$ and the induced Pr order parameter
$L^{Pr}_{3z}$ then have to point in opposite directions at the muon
site. Therefore the $\mu$SR frequency in PrFeAsO in the Fe AFM phase
is smaller compared to the one in the other compounds.

Recent neutron diffraction studies reveal that in the Pr ordered
phase the Pr moments are directed along the $c$--axis, which creates
a $L^{Pr}_{3z}$ non--zero magnetic order parameter, i.e. the same
basis function of the irreducible representation as the induced Pr
moments in the Fe ordered phase.\cite{Zhao08b} Therefore, it is
reasonable to assume that also in the Pr--ordered state, the local
field at the muon site created by the Fe and the Pr magnetic order
have opposite signs. Respective magnetic structures are shown in
Fig. \ref{img.Fig-MagMoments-above}(b). The magnetic structure shown
can therefore qualitatively explain the 2~MHz drop of the $\mu$SR
frequency in the Pr ordered phase. However, the quantitative
evaluation of the $\mu$SR frequency using (\ref{1.6}) and the Fe and
Pr moment of 0.48~\mb{}/Fe and 0.84~\mb{}/Pr at 5~K that were
deduced from neutron measurements\cite{Zhao08b} reveal a strong
discrepancy. If the suggested magnetic structure would be realized a
$\mu$SR frequency near to zero should have been observed. Even if
one includes a strong increase of the local Fe moment (neglecting
all transferred hyperfine field contributions) below the Pr ordering
temperature which might be suggested by the Mössbauer
data,\cite{McGuire08b-arXiv} the calculated $\mu$SR frequency is
much too small. To remove the discrepancy between neutron and
$\mu$SR observations one has to conclude that in the PrFeAsO the Pr
magnetic structure is either easy--plane non--collinear or
collinear. In this case the situation would be similar to the Ce and
Nd compounds where in addition to the $L^{\Pr }_{3z}(\mathbf
K_{II})$ order parameter also other $ab$--plane components like
$L^{Pr}_{3x}(\mathbf K_{II})$ or $L^{Pr}_{3y}(\mathbf K_{II})$
exist. Similar conclusions have been drawn from recent neutron
diffraction studies.\cite{Kimber08} The consequences of
non--collinear order parameters will be discussed in
Sec.~\ref{sec.non-collinear-re-re}.

\subsubsection{SmFeAsO}

Compared to La and Pr, the Sm compound demonstrates a very similar $\mu$SR response in the Fe ordered AFM phase which
implies a $L_{3x}(\mathbf K_I)$ or $L_{3x}(\mathbf K_{II})$ magnetic Fe order parameter. In spite of the similarity of
the lowest CEF levels in Sm and Ce compounds the low value of the Sm g-factor reduces the induced Sm magnetic
moment and therefore its contribution to the $\mu$SR frequency in the Fe AFM phase. While in the Sm ordered
phase, in contrast to the Pr and Ce compounds, the $\mu$SR spectra change drastically and several well resolved
frequencies are observed. At the lowest temperatures three $\mu$SR frequencies are observed: 15~MHz, 23~MHz, 31~MHz
for which the approximate ratio of amplitudes is 1:4:1. This behavior indicates that for the given magnetic
symmetry of the Sm order, at least three muon sites are inequivalent. This inequivalence can be caused by
different translational symmetry of Sm and Fe magnetic order parameters or a complex non--collinear magnetic order
of the Sm subsystem. A minimal model which can explain the $\mu$SR spectra involves a $L_{3x}(\mathbf K_I)$
order parameter for the Fe subsystem with $K_I$ and $1/2 \left[ L^{Sm}_{1z}(\mathbf K_{II}) + L^{Sm}_{3z}(\mathbf K_{II})
\right]$ and $1/\sqrt{8} \left[ L^{Sm}_{1x}(\mathbf K_{II}) - L^{Sm}_{3x}(\mathbf K_{II}) + L^{Sm}_{1y}(\mathbf K_{II}) -
L^{Sm}_{3y}(\mathbf K_{II}) \right]$ order parameters for the Sm subsystem with $K_{II}$ translational symmetries.
This magnetic structure is shown in Fig.~\ref{img.Fig-MagMoments-below}(c). It results in three inequivalent
muon sites. The first with a low local field at the muon positions (15~MHz at A5 and A7), the second where the
field from the \textit{R} system cancels (23~MHz at A2, A4, A6, and A8) and a third with a high local field (31~MHz at
A1 an A3). The experimentally observed amplitudes are also reasonably well reproduced by the model. Within this
magnetic model the size of the Sm magnetic moment can be estimated to be 0.4~$\mu_B$/Sm at 1.9~K. Note that it
is possible to fit the low temperature data with even more frequencies. The present statistical accuracy is not sufficient to determine the correctness of such a fit, but in a recent $\mu$SR study the low
temperature data have been fitted with five frequencies.\cite{DrewNature} One has to note that our symmetry
analysis also allows more complicated and even incommensurate Sm magnetic structures.\cite{Lamonova}

\section{Rare earth non--collinear order in \textit{R}FeAsO}\label{sec.non-collinear-re-re}

The above proposed \textit{R} magnetic structures for the Ce, Sm and probably also in the Pr compound involve
the well known exchange non--collinearity, i.e. a perpendicular orientation of neighboring \textit{R} moments. As shown
above, such a magnetic structure can be described by a composition of the Cartesian components of two different permutation
 modes like $L^R_{1x}(\mathbf K_{III}) + L^R_{3y}(\mathbf K_{II})$ in the case of the Ce compound or
like $1/2 \left[ L^R_{1z}(\mathbf K_{II}) + L^R_{3z}(\mathbf K_{II})
\right]$ in the case of SmFeAsO. From a pure symmetry point of view,
commensurate and non--collinear magnetic order is characteristic for
crystallographic structures with higher than orthorhombic symmetry
that implies existence of two-- or three--dimensional IR for their
space groups at k=0. From a pure energy point of view, the exchange
non--collinearity results in competition of two or three different
type of exchange multiplets (permutation modes) which belong to the
same IR and therefore possess the same Heisenberg energy. In high
symmetry magnets this competition is resolved by accounting for
fourth or higher order magnetic interactions in the system's free
energy.\cite{Sobolev93,Pashkevich95} Due to its crystallographic
structure the condition for accidental energy degeneracy of $L_1^R$
and $L_3^R$ rare earth permutation modes is fulfilled in the
\textit{R}FeAsO compounds. The part of the Hamiltonian containing
the \textit{R}--\textit{R} exchange interactions for $K_{II}$
translational symmetry can be written in the form:

\begin{equation}
\label{1.9} \mathcal{H}_{ex}^{R-R} = \ldots + J_1^R(\mathbf{L}_1^R)^2 + J_3^R(\mathbf{L}^R_3)^2 + \ldots
\end{equation}

\noindent where

\begin{equation}
\label{eqaa} J_1^R = K_{11}^{R-R} + K_{12}^{R-R} - K_{13}^{R-R} - K_{14}^{R-R}
\end{equation}

\noindent and

\begin{equation}
\label{eqbb} J_3^R = K_{11}^{R-R} - K_{12}^{R-R} - K_{13}^{R-R} + K_{14}^{R-R}.
\end{equation}

\begin{figure}[htb]
\includegraphics[width=5.6cm]{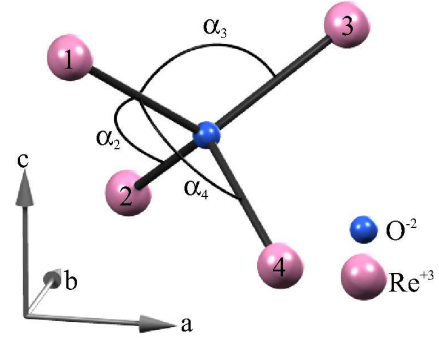}
\caption[]{The angle enumeration for \textit{R}--O--\textit{R}
magnetic bonds in the rare earth tetrahedron of \textit{R}FeAsO
(\textit{R} = Pr, Sm, Ce).} \label{img.Fig-Re-tetrahedron}
\end{figure}

\noindent For simplicity we omit the notation $K_{II}$. The symbol $K_{1\beta}^{R-R}$ denotes an exchange interaction coupling constant between nearest neighbor rare earth ions on sites 1 and $\beta$ (see Fig.~\ref{img.Fig-Re-tetrahedron}. To deduce equation
(\ref{1.9}) we used a permutation symmetry in which the interaction between ion 2 and 4 is equal to the
interaction between 1 and 3 and so on. If the exchange constant $J_\alpha^R$ is negative and less than the others $J_\beta^R$ $(\beta \ne \alpha)$ a $L_\alpha^R$ type of exchange magnetic order is realized, i.e. the relative orientation of magnetic moments (parallel/anti parallel) in the magnetic structure yields a non-zero $L_\alpha^R$ order parameter as given by equation~(\ref{1.2}).

Neutron diffraction reveals a collinear $L_3^R$-type of rare earth
exchange magnetic order in PrFeAsO ($L^{\Pr }_{3z}$ non--zero
magnetic order parameter)\cite{Zhao08b} and in NdFeAsO
(superposition of $L^{Nd}_{3x}$ and $L^{Nd}_{3z}$ non--zero magnetic
order parameters).\cite{Qiu08} Similar magnetic structures have been
proposed by a theoretical investigation of magnetic order in
\textit{R}FeAsOfor  \textit{R} = Ce, Pr.\cite{Alyahyaei09} The
authors of Ref.~\onlinecite{Alyahyaei09} claim that the \textit{R}
magnetic ground state is composed from two adjacent zigzag rare
earth chains running along the $a$--axis which carry alternating up
and a down spins. In our notation this is the $L_3^R$ type of
exchange magnetic order. However, for the Ce compound this
theoretical conclusion is not compatible with experimental data.

A possible reason for the observed non--collinearity is provided by
structural features of \textit{R}FeAsO. Note that without
orthorhombic distortion, i.e. in the tetragonal $P4/nmm$ phase, two
exchange constants $K_{12}^{R-R}$ and $K_{14}^{R-R}$ are equal to
each other, and therefore their contributions to $J_1^R$ and $J_3^R$
cancel. Moreover, even in the orthorhombic $Cmma$ phase their impact
should be small. Indeed, the sign of \textit{R}--\textit{R}
nearest-neighbor (NN) super-exchange depends on the
\textit{R}--O--\textit{R} angle. There is a critical value of this
angle at which the exchange coupling constant changes its sign. For
instance, for Cu--O--Cu bond this critical angle is equal to
104$^{\circ}$.\cite{deGraaf02} The exchange interaction is
antiferromagnetic for magnetic bonds with angles higher than this
critical angle. In the Ce compound the angles of Ce(1)--O--Ce(2) and
Ce(1)--O--Ce(4) magnetic bonds, depicted in the
Fig.~\ref{img.Fig-Re-tetrahedron} as $\alpha _2$ and $\alpha _4$,
have values of $\alpha_2=105.39^{\circ}$ and
$\alpha_4=105.85^{\circ}$ at $T=1.4$~K while the Ce(1)--O--Ce(3)
angle is $\alpha_3=117.49^{\circ}$.\cite{Zhao08a} For NdFeAsO the
respective angles are: $\alpha _2=105.13^{\circ}$, $\alpha
_4=105.58^{\circ}$, and $\alpha _3=118.06^{\circ}$ at
$T=0.3$~K.\cite{Qiu08}

The angles $\alpha_2$ and $\alpha_4$ are always very similar in
\textit{R}FeAsO. This implies that the exchange constants are also
similar even if the value is far from the critical angle. The value
is far from the critical angle that has been determined for the
Cu--O--Cu bond, but might well be near critical for the
\textit{R}--O--\textit{R} magnetic bonds considered here. The
difference between Ce(1)--Ce(2) and Ce(1)--Ce(4) distances is only
0.011~\AA.\cite{Zhao08a} Therefore, mainly antiferromagnetic
exchange $K_{13}^{R-R} > 0$ determines the magnitude of the $J_1^R$
and $J_3^R$ coupling constants. In the limit of $J_1^R = J_3^R =
J^R$ the corresponding exchange Hamiltonian is:

\begin{eqnarray}
\mathcal{H}_{ex}^{R-R} &=& J^R \left[ (\mathbf{L}_1^R)^2+(\mathbf{L}_3^R)^2 \right]\nonumber\\
& = & 1/8J^R \left[ (\mathbf{m}_1^R - \mathbf{m}_3^R)^2 + (\mathbf{m}_2^R - \mathbf{m}_4^R)^2 \right]
\label{1.10}
\end{eqnarray}

Using equation (\ref{1.10}) one can qualitatively explain the
features of \textit{R} magnetic order in the \textit{R}FeAsO
compounds. In the \textit{R}--O--\textit{R} layer both planes of the
\textit{R} ions order antiferromagnetically with non--zero
antiferromagnetic vectors $\mathbf{l}_1^R = \mathbf{m}_1^R -
\mathbf{m}_3^R$ and $\mathbf{l}_2^R = \mathbf{m}_2^R -
\mathbf{m}_4^R$ for \textit{R}--planes which are located above and
below the oxygen plane respectively. However, these two rare earth
planes are not magnetically coupled, i.e. the mutual orientation of
those vectors is not defined by a Heisenberg exchange interaction.
This magnetic frustration can be lifted by higher order interactions
like biquadratic exchange $D^R [(\mathbf{L}_1^R)^2 -
(\mathbf{L}_2^R)^2]^2 \propto D^R ( \mathbf{l}_1^R \cdot
\mathbf{l}_2^R )^2$.\cite{Sobolev93} For the case of $J^R<0<D$ this
results in a non--collinear magnetic ground state with
$\mathbf{l}_1^R \bot \mathbf{l}_2^R$. Note that this is exactly the
spin structure proposed for SmFeAsO in this study, and that it is
very similar to the one proposed by Zhao et al. \cite{Zhao08a} for
CeFeAsO which is also consistent with our data. For the case
$J^R<0$, $D<0$ a collinear structure with $\mathbf{l}_1^R \parallel
\mathbf{l}_2^R$ is obtained, this structure is realized in NdFeAsO
and probably also in PrFeAsO. The actual orientation of the
\textit{R} magnetic moments is determined by the complex hierarchy
of second order anisotropy and fourth order
interactions.\cite{Lamonova} The orthorhombic distortions observed
in the \textit{R}FeAsO system should lead to a weak coupling of
antiferromagnetic \textit{R} planes below and above the oxygen
plane. Up to now only the \textit{R}--O--\textit{R} exchange paths
have been considered. Note that an inclusion of the
\textit{R}--As--\textit{R} exchange paths does not change the
proposed scenario because the \textit{R}--As--\textit{R} magnetic
bonds only contribute to the $K_{13}^{R-R}$ exchange and not to the
\textit{R}(1)--\textit{R}(2) and \textit{R}(1)--\textit{R}(4)
exchanges. The hypothetical case where the ferromagnetic
\textit{R}(1)--As--\textit{R}(3) exchange cancels the
antiferromagnetic \textit{R}(1)--O--\textit{R}(3) exchange is very
unlikely.

\section{Influence of the \textit{R} magnetic order on the Fe sublattice}\label{sec.influence-Re-on-Fe}

In the above discussion we focused on the action of the Fe order on the \textit{R} subsystem. In accordance with
Table \ref{tab.X1} and the Hamiltonian (\ref{1.4}), the $L_{3x}$ Fe order parameter creates a molecular field on
the \textit{R} sites which is directed along z--axis and has a $L^R_{3z}$ symmetry. The reverse action of the \textit{R}
magnetic order on the Fe subsystem should also be taken into account. This action will be especially important for
a realization of the easy plane anisotropy in the \textit{R} subsystem when the rare earth and iron order
parameters possess different symmetry. The onset of the $L^R_{3x}$ or $L^R_{1y}$ in plane \textit{R} order
parameters will induce an out of plane canting of the iron moments along the z--axis due to these Fe--\textit{R}
interactions:

\begin{eqnarray}
\label{1.11}
\mathcal{H}_{an-ex}^{Fe-R} & = & \ldots + I_{3xz}^{Fe-R}(\mathbf K_{II})L_{3x}(\mathbf K_{II})L_{3z}^R(\mathbf K_{II}) + \nonumber\\
& & + I_{3zx}^{Fe-R}(\mathbf K_{II})L_{3z}(\mathbf K_{II})L_{3x}^R(\mathbf K_{II}) + \nonumber\\
& & + I_{1zy}^{Fe-R}(\mathbf K_{II})L_{1z}(\mathbf K_{II})L_{1y}^R(\mathbf K_{II}) + \nonumber\\
& & + I_{3yz}^{Fe-R}(\mathbf K_{II})L_{1y}(\mathbf K_{II})L_{1z}^R(\mathbf K_{II}) + \dots \nonumber\\
\end{eqnarray}

\begin{figure}[htb]
\includegraphics[width=8.6cm,clip]{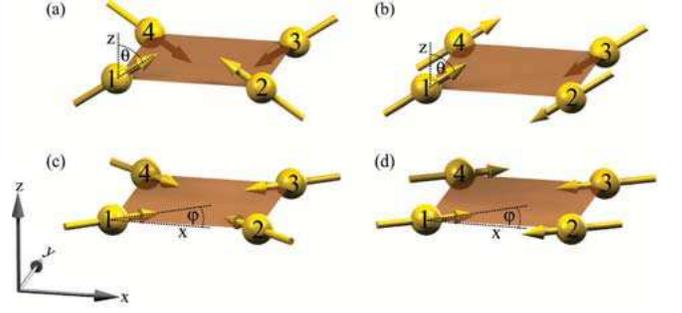}
\caption[]{Different types of canted Fe structures that are induced by different rare earth magnetic structures.
(a) Out of plane non--collinear Fe order $(L_{3x}\cos \theta + L_{1z} \sin \theta)$ induced by $L^R_{1y}$-type
rare earth order; (b) out of plane collinear Fe order $(L_{3x}\cos \theta + L_{3z} \sin \theta)$ induced by
$L^R_{3x}$-type rare earth order; (c) in plane non--collinear Fe order $(L_{3x}\cos \phi + L_{1y} \sin \phi)$
induced by $L^R_{1z}$-type rare earth order; (d) in plane collinear Fe order $(L_{3x}\cos \phi + L_{3y} \sin
\phi)$. Here, $\theta$ and $\phi$ are the polar angles between the Fe(1) magnetic moment and the z--axis and the
x--axis, respectively.} \label{img.canting}
\end{figure}

Interestingly, on the one hand the non--collinearity in the \textit{R} system ($L_{3z}^R$ and $L_{1y}^R$ order
parameter) also induces a non--collinearity in the Fe order ($L_{3x}$ and $L_{1z}$ order parameters). On the
other hand, a collinear and oblique \textit{R} magnetic order ($L^R_{3x}$ and $L^R_{3z}$ order parameters) induces
a Fe magnetic structure which is still collinear but the Fe moments deviate from the x--direction. Different types
of canted Fe structures are shown in Fig. \ref{img.canting}.

We discuss now whether a canting of the magnetic structures could be observed by $\mu$SR and Mössbauer
spectroscopy. Here we consider the contribution from the Fe sublattice only. Supposing that the size of the Fe
magnetic moment does not change below the \textit{R} ordering temperature, i.e. it is close to saturation, the $\mu$SR
response is affected by the canting in the following way: From the dipole field calculation presented in
Appendix A, it follows that the magnetic field at the muon site created by both non--collinear Fig.
\ref{img.canting}(a) and collinear Fig. \ref{img.canting}(b) out--of--plane structures is tilted by the
out--of--plane angle $\theta$ but does not change its modulus. Therefore, $\mu$SR experiments on powder samples
cannot distinguish the canted structures from the original collinear $L_{3x}$ structure. In contrast, the field
at the muon site, i.e. the $\mu$SR frequencies depend on the angle $\phi$ for in--plane canted structures shown
in Fig. \ref{img.canting}(c) and (d). In the case of the magnetic structure shown in Fig. \ref{img.canting}(d)
the $\mu$SR frequency always decreases like $\cos \phi$, whereas, in the case of structure (c) in Fig.
\ref{img.canting} the field at the muon site may decrease or increase depending on the sign of $\phi$ and has the
following angle dependence $\sqrt 2 \cos (\phi + \pi/4)$. In turn, its sign determines the sign of the coupling
constant $I_{3yz}^{Fe-R}$ of the Fe--\textit{R} interaction.

Mössbauer spectroscopy, in contrast to $\mu$SR, should be very
sensitive to the orientation of the Fe moments. Our calculation of
the electric field gradient tensor on Fe site reveals a strong
anisotropy with the largest component in the z--direction. This
makes Mössbauer spectroscopy a suitable tool to detect the direction
of the Fe magnetic moment especially in the case where out of
$ab$--plane deviations from x-axis are present. Again supposing an
almost constant modulus of the iron moments below $T_N^{R}$ one can
derive a strong increase of the quadrupole shift along with an
increase of the average hyperfine field for the out of plane canted
iron structures like those shown in Fig. \ref{img.canting}(a) and
(b). This behavior has already been detected in the Mössbauer
studies of PrFeAsO.\cite{McGuire08b-arXiv} McGuire et al. come to
the conclusion that the Fe spin-reorientation with a significant
component of Fe moment along the $c$--axis takes place below the Pr
ordering temperature. For the in-plane collinear and non--collinear
canted structures shown in Fig. \ref{img.canting}(c) and (d) the
dependence of Mössbauer spectra on the rotation angle $\phi$ should
be weaker than for out--of--plane rotation since it only changes the
asymmetry parameter of the electric field gradient tensor.

Combining both  comments above, one can understand why the drastic
change of PrFeAsO Mössbauer spectra below $T_N^{Pr}$ has no analogy
in the $\mu$SR spectra. The following model can unify both
observations:

i) In the Pr subsystem, the $L_3^R$ type of exchange order is realized with an easy plane anisotropy.

ii) The order parameter which appears below $T_N^{Pr}$ is $L_{3x}^{Pr}$ with a small induced $L_{3z}^{Pr}$
component remaining even down to the lowest temperatures due to the Fe molecular field $I_{3xz}^{Fe-Pr} (\mathbf
K_{II} )L_{3x}(\mathbf K_{II})$. The latter component being negative as mentioned above.

iii) Below $T_N^{Pr}$ the in-plane collinear Fe magnetic structure gradually rotates into the collinear
out-of-plane $L_{3x} \cos \theta +L_{3z} \sin \theta$ structure.

\noindent Respective magnetic fields at the muon sites expressed in MHz are given by

\begin{eqnarray}
 B_x(A) &=& f_0 \sin \theta - 13.55\cdot 0.59 \cdot L_{3x}^{Pr}(\mathbf K_{II})\nonumber\\
 B_z(A)&=& f_0 \cos \theta. \label{1.12}
\end{eqnarray}

\noindent Here $f_0$ is the low temperature frequency extrapolated from temperatures well above $T_N^{Pr}$ in
the Fe antiferromagnetic phase. It is determined by the saturated Fe magnetic moment and has a value of
$f_0=22$~MHz in the Pr compound. In (\ref{1.12}) the small contribution from the $L_{3z}^R$ component which
decreases as the ordered iron moment rotates is omitted. The decrease of the $\mu$SR frequency below
$T_N^{Pr}$ is roughly 2~MHz. Assuming a magnetic moment of approximately 1$\mu_B$/Pr a canting of $\sin \theta
= 0.42$ is obtained from Eq.~(\ref{1.12}). This estimation confirms indeed a strong tilting of the Fe moments to
the z--axis which was already concluded from Mössbauer spectroscopy experiments.\cite{McGuire08b-arXiv} The
respective magnetic structure is shown in Fig.~(\ref{img.Fig-MagMoments-below}b). Note, that in this model the
angle $\theta$ can achieve $90^{\circ}$, if we suppose the Pr moments equal to 0.26$\mu_B$/Pr. As mentioned
above, the $\mu$SR signal would be exactly the same for another combined Fe--Pr magnetic structure with a non--zero
non--collinear $L_{3x} \cos \theta +L_{1z} \sin \theta$ in the Fe and an $L_{1y}^{Pr}$ order parameters in the Pr
subsystem. We conclude that these two different magnetic structures are indistinguishable by $\mu$SR and
Mössbauer studies.

Finally, we comment on other aspects of the Fe--\textit{R}
interactions. It is generally known that the Fe--\textit{R}
interaction plays a decisive role for the magnetic anisotropy of the
Fe subsystem in Fe--\textit{R} compounds. For instance, temperature
induced spin reorientation phase transitions, even far above of rare
earth magnetic order temperature, have been observed in rare earth
orthoferrites ReFeO$_3$\cite{Belov76} and in the ReFe$_{11}$Ti
compounds.\cite{Piquer06} The temperature dependence of the Fe
magnetic anisotropy is determined by competing contributions from
the Fe and \textit{R} subsystems. In ReFeO$_3$ compounds the
\textit{R} ordering is almost always accompanied by a Fe spin
reorientation, and the canting angles are relatively large due to a
strong renormalization of the Fe magnetic anisotropy. In this sense
it would be surprising if the \textit{R}FeAsO system were an
exception to this rule. Furthermore it is clear that the large
proposed Fe canting angles $\theta$ in Pr ordered phase of PrFeAsO
can be achieved by a renormalization of the Fe magnetic anisotropy
but not solely by the direct Fe--Pr interaction mechanism
(\ref{1.11}).

Note that our $\mu$SR studies of \textit{R}FeAsO with Sm, Pr and Ce
do not show temperature induced spin reorientations of the iron
magnetic structure in the Fe AFM phase. Spin reorientation starts at
a temperature at which the anisotropy changes sign, i.e. it develops
as a standard second order phase transition. These have not been
detected, neither by change of the muon spin relaxation rate nor by
other methods like specific heat or susceptibility studies. In the
low temperature range the iron spin reorientation begins with the
onset of the rare earth magnetic order and therefore it can be
masked by the latter.

\section{Conclusion}
In conclusion we presented a detailed study of the magnetic order
and the interplay of the rare earth and iron magnetism in
\textit{R}FeAsO with \textit{R} = La, Ce, Pr, and Sm, the
magnetically ordered parent compounds of the recently discovered
\textit{R}O$_{1-x}$F$_x$FeAs high temperature superconductors. Using
zero field (ZF) $\mu$SR, the N\'eel temperatures, as well as the
temperature dependence of the sublattice magnetizations, have been
determined with high precision.

The second order Fe magnetic phase transition is well--separated
from a structural transition which occurs 10--20~K above $T_{N}$.
The muon site in the \textit{R}FeAsO crystal structure has been
obtained by electronic potential calculations involving a modified
Thomas--Fermi approach. Using calculated dipole fields at the muon
site, our $\mu$SR experiments indicate an antiferromagnetic
commensurate order with the iron magnetic moments directed along the
$a$-axis above $T_N^{R}$. This is consistent with neutron scattering
results reported earlier.\cite{Lynn08} The calculations show, that
the two types of iron order parameter with different translational
symmetry, that have been observed in \textit{R}FeAsO compounds, i.e.
with ($\mathbf K_{I}$) and without ($\mathbf K_{II}$) doubling of
the magnetic primitive cell along the $c$--direction cause nearly
the same field at the muon site and are therefore not
distinguishable by $\mu$SR. Mössbauer spectroscopy
\cite{Klauss08,McGuire08b-arXiv} as well as $\mu$SR prove that the
ordered Fe magnetic moment is approximately 0.4~\mb{} and does not
vary by more than 15~\% within the series \textit{R}FeAsO with
\textit{R} = La, Ce, Pr, and Sm. This is in sharp contrast to
published neutron results \cite{Lynn08} where Fe magnetic moments of
0.36~\mb{}, 0.48~\mb{}, and 0.8~\mb{} have been deduced for
\textit{R} = La, Pr, and Ce respectively. In the neutron studies,
the observed intensity in the magnetic Bragg peaks is solely
attributed to an ordered Fe moment for temperatures larger than
$T_N^{R}$. Here we could show that a sizable magnetic moment of the
\textit{R} subsystem which is induced by the ordered Fe subsystem is
already detectable far above $T_N^{R}$. Especially in CeFeAsO this
\textit{R} magnetization is exceptionally large and amounts to
approximately 0.3~\mb{}/Ce just above $T_N^{Ce}$. Since for symmetry
reasons the Ce and Fe magnetic moments contribute to the same Bragg
peaks, neglecting the Ce magnetization results in the strong
overestimation of the Fe moment in the CeFeAsO system by neutron
scattering.

Static magnetic ordering of the rare earth moments has been observed
for the \textit{R} = Ce, Pr, and Sm compounds with transition
temperatures of $T_{N}^{R}$~=~4.4(3), 11(1), and 4.66(5)~K
respectively. Using available literature data of Mössbauer
spectroscopy and neutron diffraction and the $\mu$SR data presented
here, we propose combined \textit{R} and Fe magnetic structures
below $T_{N}^{R}$ for all investigated compounds. For CeFeAsO, the
$\mu$SR data are consistent with a non--collinear easy plane AFM
order of the Ce moments as concluded from neutron
diffraction.\cite{Zhao08a} For SmFeAsO, several new $\mu$SR
frequencies develop in the Sm ordered phase. To explain the $\mu$SR
spectra a minimal model of non--collinear Sm magnetic order is
proposed. The additional $\mu$SR frequencies originate from
different translational symmetries of the Sm and Fe order
parameters. Analyzing structural features of \textit{R}FeAsO we
argue that the non--collinear rare earth magnetic order observed in
Ce and Sm compounds arises due to a weak magnetic coupling of the
adjacent \textit{R} planes in the \textit{R}--O--\textit{R} layer. A
weak coupling of the \textit{R}--O--\textit{R} layers is probably an
inherent feature of all \textit{R}FeAsO compounds. For PrFeAsO, we
propose that the Pr moments order also with an easy plane
anisotropy. Our model therefore differs from the one determined by
Zhao et al. \cite{Zhao08b} which implies an easy axis anisotropy in
the Pr ordered phase. Only the model with easy plane anisotropy is
able to consistently describe the $\mu$SR and Mössbauer
\cite{McGuire08b-arXiv} as well as the neutron data.

Even though Fe spin reorientation phase transitions are frequently
observed in \textit{R}--Fe systems, e.g. in rare earth orthoferrites
\textit{R}FeO$_3$\cite{Belov76} or \textit{R}Fe$_{11}$Ti
\cite{Piquer06}, we do not find such a transition in \textit{R}FeAsO
above $T_{N}^{R}$ for \textit{R} = Ce, Pr and Sm. In the \textit{R}
ordered phase one can expect a strong influence of the rare earth
magnetic subsystem on the Fe ordering direction especially in the
case when the iron and rare earth order parameters have different
symmetry. This is the case for collinear as well as non--collinear
easy plane  AFM order in the \textit{R} subsystem. We suggest that
this is the reason for the Fe spin reorientation developing below
$T_{N}^{Pr}$ as deduced from the $\mu$SR data on PrFeAsO.

Using symmetry arguments we demonstrate the absence of a Heisenberg
magnetic interaction between the Fe and the \textit{R} subsystem in
\textit{R}FeAsO. Therefore, the apparent Fe--\textit{R} interaction
is realized by a non--Heisenberg anisotropic exchange. Our
calculations additionally show that dipole--dipole interactions are
much too weak to account for the observed couplings. We showed that
in CeFeAsO the Fe--Ce coupling is exceptionally large. This together
with the large paramagnetic Ce magnetic moment explains the sizable
Ce magnetization observed in our experiment in the Fe AFM ordered
phase as well as the presence of a small component of Ce moment
along the $c$--axis in the Ce AFM ordered phase. For PrFeAsO, from a
comparison of $\mu$SR, Mössbauer and neutron diffraction data we
derive a much weaker but noticeable Fe--Pr coupling constant which
has opposite sign compared to the Fe--Ce coupling constant. For
SmFeAsO the Fe-Sm coupling is very weak and the induced polarization
of the Sm moments by the magnetically ordered Fe system is not
measurable above $T_{N}^{Sm}$. Even though the lowest CEF levels in
Sm and Ce are relatively similar, the low value of the Sm g-factor
reduces the induced Sm magnetic moment and therefore its
contribution to the $\mu$SR frequency. The exceptionally strong
coupling of the Ce to the Fe subsystem in CeFeAsO can also be
understood from band structure calculations: Only CeFeAsO possesses
a considerable 3d--4f
hybridization.\cite{Vildosola08,Pourovskii08,Miyake08} Additionally,
the strong interaction of the two magnetic sublattices is in
accordance with the observed strong electron correlation of the Ce
4f electrons and the heavy fermion behavior observed in the related
phosphide CeOFeP. \cite{Bruening08}

Finally, from our analysis we can conclude that the magnetic
\textit{R}--Fe interaction is probably not crucial for the observed
enhanced superconductivity in \textit{R}O$_{1-x}$F$_x$FeAs with
magnetic 4f ions compared to LaO$_{1-x}$F$_x$FeAs, since only in
CeFeAsO a strong \textit{R}--Fe coupling is observed, while it is
much weaker in the Pr and Sm compounds.

\begin{acknowledgments}
We would like to thank M. Zhitomirsky, R. Moessner, I.M. Vitebsky
and N.M. Plakida for useful discussions. We thank M. Deutschmann, S.
M\"uller--Litvanyi, R. M\"uller and A. K\"ohler for experimental
support in preparation and characterization of the samples. The work
at TU Dresden has been supported by the DFG under Grant No.
KL1086/8-1. Yu.G.P. acknowledges the support of the
Ukrainian-Russian grant 2008--8. The work at the IFW Dresden has
been supported by the DFG through FOR 538. Part of this work was
performed at the Swiss Muon Source (Villigen, Switzerland).
\end{acknowledgments}

\section{Appendix}
\label{sec.app}
\subsection{Magnitude and symmetry of dipole fields created by the iron and rare earth
subsystems at the A type muon site}

Below we provide the calculation of magnetic dipole fields at the muon site A for different magnetic modes and
different translational symmetries. The magnetic fields below are given in units of 10$^{-1}$ T and basis
functions $L$ in units of~\mb{}. The
coordinates of the muon site A are (0,1/4,0.41)

\begin{widetext}
\begin{flushleft}
\begin{equation*}
\label{eq7}
\begin{array}{l}
 \left( {{\begin{array}{*{20}c}
 {B_x (A )} \hfill \\
 {B_y (A )} \hfill \\
 {B_z (A )} \hfill \\
\end{array} }} \right)=\left( {{\begin{array}{*{20}c}
 0 \hfill & 0 \hfill & 0 \hfill \\
 0 \hfill & 0 \hfill & {-2.56} \hfill \\
 0 \hfill & {-2.56} \hfill & 0 \hfill \\
\end{array} }} \right)\left( {{\begin{array}{*{20}c}
 {L_{1x} (\mathbf K_I )} \hfill \\
 {L_{1y} (\mathbf K_I )} \hfill \\
 {L_{1z} (\mathbf K_I )} \hfill \\
\end{array} }} \right)+\left( {{\begin{array}{*{20}c}
 0 \hfill & 0 \hfill & {2.58} \hfill \\
 0 \hfill & 0 \hfill & 0 \hfill \\
 {2.58} \hfill & 0 \hfill & 0 \hfill \\
\end{array} }} \right)\left( {{\begin{array}{*{20}c}
 {L_{3x} (\mathbf K_I )} \hfill \\
 {L_{3y} (\mathbf K_I )} \hfill \\
 {L_{3z} (\mathbf K_I )} \hfill \\
\end{array} }} \right)+ \\ \\ \left( {{\begin{array}{*{20}c}
 {-0.79} \hfill & 0 \hfill & 0 \hfill \\
 0 \hfill & {-0.66} \hfill & 0 \hfill \\
 0 \hfill & 0 \hfill & {1.45} \hfill \\
\end{array} }} \right)\left( {{\begin{array}{*{20}c}
 {L^R_{1x} (\mathbf K_I )} \hfill \\
 {L^R_{1y} (\mathbf K_I )} \hfill \\
 {L^R_{1z} (\mathbf K_I )} \hfill \\
\end{array} }} \right)+\left( {{\begin{array}{*{20}c}
 {-0.66} \hfill & 0 \hfill & 0 \hfill \\
 0 \hfill & {-0.79} \hfill & 0 \hfill \\
 0 \hfill & 0 \hfill & {1.45} \hfill \\
\end{array} }} \right)\left( {{\begin{array}{*{20}c}
 {L^R_{3x} (\mathbf K_I )} \hfill \\
 {L^R_{3y} (\mathbf K_I )} \hfill \\
 {L^R_{3z} (\mathbf K_I )} \hfill \\
\end{array} }} \right) \\ \end{array}
\end{equation*}

\begin{equation*}
\label{eq8}
\begin{array}{l}
 \left( {{\begin{array}{*{20}c}
 {B_x (A )} \hfill \\
 {B_y (A )} \hfill \\
 {B_z (A )} \hfill \\
\end{array} }} \right)=\left( {{\begin{array}{*{20}c}
 0 \hfill & 0 \hfill & 0 \hfill \\
 0 \hfill & 0 \hfill & {-2.55} \hfill \\
 0 \hfill & {-2.55} \hfill & 0 \hfill \\
\end{array} }} \right)\left( {{\begin{array}{*{20}c}
 {L_{1x} (\mathbf K_{II} )} \hfill \\
 {L_{1y} (\mathbf K_{II} )} \hfill \\
 {L_{1z} (\mathbf K_{II} )} \hfill \\
\end{array} }} \right)+\left( {{\begin{array}{*{20}c}
 0 \hfill & 0 \hfill & {2.57} \hfill \\
 0 \hfill & 0 \hfill & 0 \hfill \\
 {2.57} \hfill & 0 \hfill & 0 \hfill \\
\end{array} }} \right)\left( {{\begin{array}{*{20}c}
 {L_{3x} (\mathbf K_{II} )} \hfill \\
 {L_{3y} (\mathbf K_{II} )} \hfill \\
 {L_{3z} (\mathbf K_{II} )} \hfill \\
\end{array} }} \right)+ \\ \\ \left( {{\begin{array}{*{20}c}
 {-0.89} \hfill & 0 \hfill & 0 \hfill \\
 0 \hfill & {-0.59} \hfill & 0 \hfill \\
 0 \hfill & 0 \hfill & {1.48} \hfill \\
\end{array} }} \right)\left( {{\begin{array}{*{20}c}
 {L^R_{1x} (\mathbf K_{II} )} \hfill \\
 {L^R_{1y} (\mathbf K_{II} )} \hfill \\
 {L^R_{1z} (\mathbf K_{II} )} \hfill \\
\end{array} }} \right)+\left( {{\begin{array}{*{20}c}
 {-0.59} \hfill & 0 \hfill & 0 \hfill \\
 0 \hfill & {-0.88} \hfill & 0 \hfill \\
 0 \hfill & 0 \hfill & {1.48} \hfill \\
\end{array} }} \right)\left( {{\begin{array}{*{20}c}
 {L^R_{3x} (\mathbf K_{II} )} \hfill \\
 {L^R_{3y} (\mathbf K_{II} )} \hfill \\
 {L^R_{3z} (\mathbf K_{II} )} \hfill \\
\end{array} }} \right) \\ \end{array}
\end{equation*}

\begin{equation*}
\label{eq9}
\begin{array}{l}
 \left( {{\begin{array}{*{20}c}
 {B_x (A )} \hfill \\
 {B_y (A )} \hfill \\
 {B_z (A )} \hfill \\
\end{array} }} \right)=\left( {{\begin{array}{*{20}c}
 {2.16} \hfill & 0 \hfill & 0 \hfill \\
 0 \hfill & {2.12} \hfill & 0 \hfill \\
 0 \hfill & 0 \hfill & {-4.28} \hfill \\
\end{array} }} \right)\left( {{\begin{array}{*{20}c}
 {L_{1x} (\mathbf K_{III} )} \hfill \\
 {L_{1y} (\mathbf K_{III} )} \hfill \\
 {L_{1z} (\mathbf K_{III} )} \hfill \\
\end{array} }} \right)+\left( {{\begin{array}{*{20}c}
 0 \hfill & {-4.44} \hfill & 0 \hfill \\
 {-4.44} \hfill & 0 \hfill & 0 \hfill \\
 0 \hfill & 0 \hfill & 0 \hfill \\
\end{array} }} \right)\left( {{\begin{array}{*{20}c}
 {L_{3x} (\mathbf K_{III} )} \hfill \\
 {L_{3y} (\mathbf K_{III} )} \hfill \\
 {L_{3z} (\mathbf K_{III} )} \hfill \\
\end{array} }} \right)+ \\ \\ \left( {{\begin{array}{*{20}c}
 {-0.40} \hfill & 0 \hfill & 0 \hfill \\
 0 \hfill & {-0.40} \hfill & 0 \hfill \\
 0 \hfill & 0 \hfill & {0.80} \hfill \\
\end{array} }} \right)\left( {{\begin{array}{*{20}c}
 {L^R_{1x} (\mathbf K_{III} )} \hfill \\
 {L^R_{1y} (\mathbf K_{III} )} \hfill \\
 {L^R_{1z} (\mathbf K_{III} )} \hfill \\
\end{array} }} \right)+\left( {{\begin{array}{*{20}c}
 {-0.43} \hfill & 0 \hfill & 0 \hfill \\
 0 \hfill & {-0.43} \hfill & 0 \hfill \\
 0 \hfill & 0 \hfill & {0.86} \hfill \\
\end{array} }} \right)\left( {{\begin{array}{*{20}c}
 {L^R_{3x} (\mathbf K_{III} )} \hfill \\
 {L^R_{3y} (\mathbf K_{III} )} \hfill \\
 {L^R_{3z} (\mathbf K_{III} )} \hfill \\
\end{array} }} \right) \\ \end{array}
\end{equation*}

\begin{equation*}
\label{eq10}
\begin{array}{l}
 \left( {{\begin{array}{*{20}c}
 {B_x (A )} \hfill \\
 {B_y (A )} \hfill \\
 {B_z (A )} \hfill \\
\end{array} }} \right)=\left( {{\begin{array}{*{20}c}
 {1.59} \hfill & 0 \hfill & 0 \hfill \\
 0 \hfill & {1.55} \hfill & 0 \hfill \\
 0 \hfill & 0 \hfill & {-3.15} \hfill \\
\end{array} }} \right)\left( {{\begin{array}{*{20}c}
 {F_x (\mathbf K_0 )} \hfill \\
 {F_y (\mathbf K_0 )} \hfill \\
 {F_z (\mathbf K_0 )} \hfill \\
\end{array} }} \right)+\left( {{\begin{array}{*{20}c}
 0 \hfill & {-4.45} \hfill & 0 \hfill \\
 {-4.45} \hfill & 0 \hfill & 0 \hfill \\
 0 \hfill & 0 \hfill & 0 \hfill \\
\end{array} }} \right)\left( {{\begin{array}{*{20}c}
 {L_{2x} (\mathbf K_0 )} \hfill \\
 {L_{2y} (\mathbf K_0 )} \hfill \\
 {L_{2z} (\mathbf K_0 )} \hfill \\
\end{array} }} \right)+ \\ \\ \left( {{\begin{array}{*{20}c}
 {-0.95} \hfill & 0 \hfill & 0 \hfill \\
 0 \hfill & {-0.95} \hfill & 0 \hfill \\
 0 \hfill & 0 \hfill & {1.90} \hfill \\
\end{array} }} \right)\left( {{\begin{array}{*{20}c}
 {F^R_x (\mathbf K_0 )} \hfill \\
 {F^R_y (\mathbf K_0 )} \hfill \\
 {F^R_z (\mathbf K_0 )} \hfill \\
\end{array} }} \right)+\left( {{\begin{array}{*{20}c}
 {-0.44} \hfill & 0 \hfill & 0 \hfill \\
 0 \hfill & {-0.44} \hfill & 0 \hfill \\
 0 \hfill & 0 \hfill & {0.88} \hfill \\
\end{array} }} \right)\left( {{\begin{array}{*{20}c}
 {L^R_{2x} (\mathbf K_0 )} \hfill \\
 {L^R_{2y} (\mathbf K_0 )} \hfill \\
 {L^R_{2z} (\mathbf K_0 )} \hfill \\
\end{array} }} \right) \\ \end{array}
\end{equation*}
\end{flushleft}
\normalsize
\end{widetext}

\subsection{Dipole fields created by the iron and rare earth subsystems at the rare earth site}

Below we provide the calculation of magnetic dipole fields at the rare earth site for different magnetic modes and
different translational symmetries. The magnetic fields below are given in units of 10$^{-1}$ T and basis
functions $L$ in units of~\mb{}.

\begin{widetext}
\begin{flushleft}
\begin{equation*}
\label{eq11}
\begin{array}{l}
 \left( {{\begin{array}{*{20}c}
 {B_x (R)} \hfill \\
 {B_y (R)} \hfill \\
 {B_z (R)} \hfill \\
\end{array} }} \right)=\left( {{\begin{array}{*{20}c}
 0 \hfill & 0 \hfill & 0 \hfill \\
 0 \hfill & 0 \hfill & {\mbox{-0.52}} \hfill \\
 0 \hfill & {\mbox{-0.52}} \hfill & 0 \hfill \\
\end{array} }} \right)\left( {{\begin{array}{*{20}c}
 {L_{1x} (\mathbf K_I )} \hfill \\
 {L_{1y} (\mathbf K_I )} \hfill \\
 {L_{1z} (\mathbf K_I )} \hfill \\
\end{array} }} \right)+\left( {{\begin{array}{*{20}c}
 0 \hfill & 0 \hfill & {\mbox{0.53}} \hfill \\
 0 \hfill & 0 \hfill & 0 \hfill \\
 {\mbox{0.53}} \hfill & 0 \hfill & 0 \hfill \\
\end{array} }} \right)\left( {{\begin{array}{*{20}c}
 {L_{3x} (\mathbf K_I )} \hfill \\
 {L_{3y} (\mathbf K_I )} \hfill \\
 {L_{3z} (\mathbf K_I )} \hfill \\
\end{array} }} \right)+ \\ \\ +\left( {{\begin{array}{*{20}c}
 {\mbox{-0.30}} \hfill & 0 \hfill & 0 \hfill \\
 0 \hfill & {\mbox{-1.29}} \hfill & 0 \hfill \\
 0 \hfill & 0 \hfill & {\mbox{1.59}} \hfill \\
\end{array} }} \right)\left( {{\begin{array}{*{20}c}
 {L^R_{1x} (\mathbf K_I )} \hfill \\
 {L^R_{1y} (\mathbf K_I )} \hfill \\
 {L^R_{1z} (\mathbf K_I )} \hfill \\
\end{array} }} \right)+\left( {{\begin{array}{*{20}c}
 {\mbox{-1.30}} \hfill & 0 \hfill & 0 \hfill \\
 0 \hfill & {\mbox{-0.29}} \hfill & 0 \hfill \\
 0 \hfill & 0 \hfill & {\mbox{1.59}} \hfill \\
\end{array} }} \right)\left( {{\begin{array}{*{20}c}
 {L^R_{3x} (\mathbf K_I )} \hfill \\
 {L^R_{3y} (\mathbf K_I )} \hfill \\
 {L^R_{3z} (\mathbf K_I )} \hfill \\
\end{array} }} \right) \\ \end{array}
\end{equation*}

\begin{equation*}
\label{eq12}
\begin{array}{l}
 \left( {{\begin{array}{*{20}c}
 {B_x (R)} \hfill \\
 {B_y (R)} \hfill \\
 {B_z (R)} \hfill \\
\end{array} }} \right)=\left( {{\begin{array}{*{20}c}
 0 \hfill & 0 \hfill & 0 \hfill \\
 0 \hfill & 0 \hfill & {\mbox{-0.45}} \hfill \\
 0 \hfill & {\mbox{-0.45}} \hfill & 0 \hfill \\
\end{array} }} \right)\left( {{\begin{array}{*{20}c}
 {L_{1x} (\mathbf K_{II} )} \hfill \\
 {L_{1y} (\mathbf K_{II} )} \hfill \\
 {L_{1z} (\mathbf K_{II} )} \hfill \\
\end{array} }} \right)+\left( {{\begin{array}{*{20}c}
 0 \hfill & 0 \hfill & {\mbox{0.46}} \hfill \\
 0 \hfill & 0 \hfill & 0 \hfill \\
 {\mbox{0.46}} \hfill & 0 \hfill & 0 \hfill \\
\end{array} }} \right)\left( {{\begin{array}{*{20}c}
 {L_{3x} (\mathbf K_{II} )} \hfill \\
 {L_{3y} (\mathbf K_{II} )} \hfill \\
 {L_{3z} (\mathbf K_{II} )} \hfill \\
\end{array} }} \right)+ \\ \\ \left( {{\begin{array}{*{20}c}
 {\mbox{-1.32}} \hfill & 0 \hfill & 0 \hfill \\
 0 \hfill & {\mbox{ -0.28}} \hfill & 0 \hfill \\
 0 \hfill & 0 \hfill & {\mbox{1.60}} \hfill \\
\end{array} }} \right)\left( {{\begin{array}{*{20}c}
 {L^R_{1x} (\mathbf K_{II} )} \hfill \\
 {L^R_{1y} (\mathbf K_{II} )} \hfill \\
 {L^R_{1z} (\mathbf K_{II} )} \hfill \\
\end{array} }} \right)+\left( {{\begin{array}{*{20}c}
 {-0.29} \hfill & 0 \hfill & 0 \hfill \\
 0 \hfill & {\mbox{-1.30}} \hfill & 0 \hfill \\
 0 \hfill & 0 \hfill & {\mbox{1.59}} \hfill \\
\end{array} }} \right)\left( {{\begin{array}{*{20}c}
 {L^R_{3x} (\mathbf K_{II} )} \hfill \\
 {L^R_{3y} (\mathbf K_{II} )} \hfill \\
 {L^R_{3z} (\mathbf K_{II} )} \hfill \\
\end{array} }} \right) \\ \end{array}
\end{equation*}

\begin{equation*}
\label{eq13}
\begin{array}{l}
 \left( {{\begin{array}{*{20}c}
 {B_x (R)} \hfill \\
 {B_y (R)} \hfill \\
 {B_z (R)} \hfill \\
\end{array} }} \right)=\left( {{\begin{array}{*{20}c}
 {\mbox{0.03}} \hfill & 0 \hfill & 0 \hfill \\
 0 \hfill & {0.03} \hfill & 0 \hfill \\
 0 \hfill & 0 \hfill & {\mbox{-0.06}} \hfill \\
\end{array} }} \right)\left( {{\begin{array}{*{20}c}
 {L_{1x} (\mathbf K_{III} )} \hfill \\
 {L_{1y} (\mathbf K_{III} )} \hfill \\
 {L_{1z} (\mathbf K_{III} )} \hfill \\
\end{array} }} \right)+\left( {{\begin{array}{*{20}c}
 0 \hfill & {\mbox{-0.17}} \hfill & 0 \hfill \\
 {\mbox{-0.17}} \hfill & 0 \hfill & 0 \hfill \\
 0 \hfill & 0 \hfill & 0 \hfill \\
\end{array} }} \right)\left( {{\begin{array}{*{20}c}
 {L_{3x} (\mathbf K_{III} )} \hfill \\
 {L_{3y} (\mathbf K_{III} )} \hfill \\
 {L_{3z} (\mathbf K_{III} )} \hfill \\
\end{array} }} \right)+ \\ \\ \left( {{\begin{array}{*{20}c}
 {-0.12} \hfill & 0 \hfill & 0 \hfill \\
 0 \hfill & {-0.13} \hfill & 0 \hfill \\
 0 \hfill & 0 \hfill & {0.25} \hfill \\
\end{array} }} \right)\left( {{\begin{array}{*{20}c}
 {L^R_{1x} (\mathbf K_{III} )} \hfill \\
 {L^R_{1y} (\mathbf K_{III} )} \hfill \\
 {L^R_{1z} (\mathbf K_{III} )} \hfill \\
\end{array} }} \right)+\left( {{\begin{array}{*{20}c}
 {0.23} \hfill & 0 \hfill & 0 \hfill \\
 0 \hfill & {0.23} \hfill & 0 \hfill \\
 0 \hfill & 0 \hfill & {\mbox{-0.46}} \hfill \\
\end{array} }} \right)\left( {{\begin{array}{*{20}c}
 {L^R_{3x} (\mathbf K_{III} )} \hfill \\
 {L^R_{3y} (\mathbf K_{III} )} \hfill \\
 {L^R_{3z} (\mathbf K_{III} )} \hfill \\
\end{array} }} \right) \\ \end{array}
\end{equation*}
\end{flushleft}
\end{widetext}

\end{document}